\documentclass[preprint,groupedaddress,tightenlines,nofootinbib,floatfix,dvitex]{revtex4-1}

\usepackage{graphicx}
\usepackage{dcolumn}% Align table columns on decimal point
\usepackage{amssymb}
\usepackage{amsmath}
\usepackage{epstopdf}
\usepackage{xcolor}         %% for comments

\makeatletter
\newcommand{\figcaption}[1]{\def\@captype{figure}\caption{#1}}
\newcommand{\tblcaption}[1]{\def\@captype{table}\caption{#1}}
\makeatother

\def\simge{\mathrel{%
       \rlap{\raise 0.511ex \hbox{$>$}}{\lower 0.511ex \hbox{$\sim$}}}}
\def\simle{\mathrel{%
       \rlap{\raise 0.511ex \hbox{$<$}}{\lower 0.511ex \hbox{$\sim$}}}}

\begin{document}

\title{Gradient flow, confinement, and magnetic monopole in $U(1)$ lattice gauge theory}

\author{Shinji Ejiri$^{1}$ and Yuya Horikoshi$^{2}$}

\affiliation{
$^{1}$Department of Physics, Niigata University, Niigata 950-2181, Japan \\
$^{2}$Graduate School of Science and Technology, Niigata University, Niigata 950-2181, Japan}

%\date{\today}
%\date{March 31, 2023}
\date{July 23, 2023}

\begin{abstract}
In the gradient flow method of lattice gauge theory, coarse graining is performed so as to reduce the action, and as the coarse graining progresses, the field strength becomes very small.
However, the confinement property that particles interact strongly is not lost by the gradient flow.
It is seemingly mysterious, and something stable against coarse graining is expected to be behind the nature of confinement.
By performing Monte Carlo simulations of $U(1)$ lattice gauge theory, we discuss the relationship between the gradient flow and magnetic monopoles created by the compactness of the $U(1)$ gauge group.
Many magnetic monopoles are generated in the confinement phase but not so many in the deconfinement phase.
Since the monopole is a kind of topological quantity, the number of monopoles does not change much by the coarse graining.
To investigate why the confinement properties are not lost by the gradient flow, we computed Wilson loops and Polyakov loops separating them into the field strength and the monopole contributions.
We found that the field strength, which decreases with the gradient flow, does not affect confinement properties, and the monopole and the confinement properties are strongly related.
Furthermore, we discuss the relationship between the magnetic monopole and the center symmetry, which is the symmetry broken by the confinement phase transition.
\end{abstract}

\maketitle

%-------------------------------------------------------------------
\section{Introduction}
\label{sec:intro}

The gradient flow method \cite{Narayanan:2006rf,Luscher:2009eq,Luscher:2010iy} is known to be a powerful method in the study of lattice gauge theory.
Using the small flow time expansion (SF$t$X) method \cite{Suzuki:2013gza,Makino:2014taa} based on the gradient flow, the temperature dependence of thermodynamic quantities near the deconfinement phase transition point can be computed accurately \cite{Asakawa:2013laa,Kitazawa:2016dsl,Iritani:2018idk,Taniguchi:2016ofw,Taniguchi:2016tjc,Taniguchi:2020mgg}.
Moreover, the computation of the latent heat at the first-order phase transition point is possible \cite{Shirogane:2020muc}.
In the SF$t$X method, the original thermodynamic quantity is calculated using the operator after coarse graining by the gradient flow.
In order for such calculations about the confinement phase transition to be possible, it is necessary that the confinement properties must not change before and after the gradient flow.
Coarse graining is performed so as to reduce the action in the gradient flow method, and as coarse graining progresses, the strength of the gauge field $F_{\mu \nu}$ becomes very small.
Nevertheless, the fact that the important properties before flowing are not lost means that something other than the field strength gives the system its important properties.

In this paper, we consider a $U(1)$ gauge theory instead of $SU(3)$ for simplicity.
The $U(1)$ gauge theory is different from $SU(3)$ because it is not an asymptotic free theory, and the determination of lattice spacing is also ambiguous.
However, we consider that the difference between gauge groups is not so important for the change of gauge field in the process of coarse graining by the gradient flow.
In particular, we discuss the relationship between magnetic monopoles and the gradient flow in compact $U(1)$ lattice gauge theory.
In the $U(1)$ lattice gauge theory, a magnetic monopole is defined due to the compactness of the gauge group.
It is also known that the magnetic monopole gives rise to a linear potential between particles \cite{Banks:1977cc}.
Since the magnetic monopole is a topological quantity, it can be expected to be stable during coarse graining.
We consider its stability and the nature of confinement.

Topological quantities are known to play important roles also in QCD \cite{Shuryak2021}.
Abelian projection makes QCD a $U(1)$-like theory, and magnetic monopoles can be defined \cite{tHooft:1981bkw}.
Many studies have been done on the relationship between color magnetic monopoles and the quark confinement after the Abelian projection in QCD \cite{Kronfeld:1987ri,Kronfeld:1987vd,Shiba:1994ab,Ejiri:1994uw,Suzuki:1994ay,Ejiri:1995gd,DAlessandro:2007lae,Suzuki:2009xy}.
We believe that discussing the monopoles in the $U(1)$ lattice gauge theory is also helpful for understanding the color confinement in QCD as an early step.

Similar to topological quantities, symmetries are stable against coarse graining.
The finite temperature deconfinement phase transition can be understood as a spontaneous breaking of the center symmetry.
It is important to clarify the relationship among the center symmetry, its order parameter Polyakov loop, and magnetic monopoles and discuss how their properties change with the gradient flow.

In the next section, we will discuss the gradient flow in the $U(1)$ lattice gauge theory. 
In Sec.~\ref{sec:monopole}, we discuss the magnetic monopole caused by the compactness in the $U(1)$ lattice gauge theory. 
We investigate the behavior of the magnetic monopole when smearing the gauge field by the gradient flow.
Furthermore, in Sec.~\ref{sec:noncompact}, we introduce the definitions of two types of flow equations.
Then, we point out that the compactness of the gauge field is very important for having the nature of confinement.
If the gradient flow is performed without considering the compactness, the confinement property is lost and the magnetic monopole disappears.
In Sec.~\ref{sec:string}, we calculate Wilson loop and discuss the origin of the linear potential.
We decompose the Wilson loop into contributions from field strength and from monopoles, and consider their relationship with the gradient flow.
The deconfinement phase transition is discussed in Sec.~\ref{sec:transition}, focusing on the Polyakov loop.
In Sec.~\ref{sec:center}, we discuss the relationship between the gradient flow and the center symmetry.
The contribution from monopoles to the Polyakov loop has the center symmetry.
We comment on the importance of keeping the center symmetry in the gradient flow process.
Section~\ref{conclusion} provides the conclusions and future prospects of this paper.

\section{Gradient flow and confinement}
\label{sec:gradient}

In the gradient flow method of QCD proposed in Ref.~\cite{Luscher:2010iy}, the ``flowed'' gauge field $B_{\mu}^a(t,x)$ at flow time $t$ is obtained by solving the flow equation 
\begin{equation}
   \frac{\partial B_\mu^a}{\partial t} (t,x)=D_\nu G_{\nu\mu}^a(t,x) \equiv
\partial_\nu G_{\nu\mu}^a(t,x)+f^{abc}B_\nu^b(t,x)G_{\nu\mu}^c(t,x) ,
\label{eq:flow}
\end{equation}
for quenched QCD with the initial condition $B_{\mu}^a(0,x)=A_{\mu}^a(x)$, where $G_{\mu\nu}^a(t,x)$
%\begin{equation}
%G_{\mu\nu}^a(t,x) \equiv \partial_\mu B_\nu^a(t,x)-\partial_\nu B_\mu^a(t,x)+f^{abc}%B_\mu^b(t,x)B_\nu^c(t,x)
%\end{equation}
is the flowed field strength given from $B_\mu^a(t,x)$.
Because Eq.~(\ref{eq:flow}) is a kind of diffusion equation, we can regard $B_{\mu}^a(t,x)$ as a smeared field of the original gauge field $A_{\mu}^a(x)$ over a physical range of $\sqrt{8t}$ in four dimensions.
Furthermore, $-1$ times the right-hand side is equal to the functional derivative of the (flowed) action $S_g$ with respect to $B_{\mu}^a(t,x)$, $\delta S_g/ \delta B_{\mu}^a$.
Thus, the flowed field strength is weakened by the gradient flow.
Operators constructed from $B_{\mu}^a(t,x)$ (flowed operators) have no ultraviolet divergences nor short-distance singularities at finite and positive $t$, and the gradient flow defines a kind of renormalization scheme.
This method is formulated nonperturbatively and interesting results from lattice QCD calculations have been obtained so far.

However, it is mysterious that the properties of the original field are kept and the original physical quantities can be calculated from the flowed field even though the flowed field strength is weakened by the gradient flow.
To clarify why the gradient flow method works so well, we consider (compact) $U(1)$ lattice gauge theory for simplicity.
In this study, we do not deal with dynamical fermions.

The action of the gauge field is given by 
\begin{eqnarray}
S_g & = & -\beta \sum_{x, \, \mu > \nu}
       \mathrm{Re} \left[ U_{\mu}(x) U_{\nu}(x+\hat{\mu})
         U^*_{\mu}(x+\hat{\nu}) U^*_{\nu}(x) \right] \nonumber \\
   & = & -\beta \sum_{x, \mu > \nu} \cos \Theta_{ \mu \nu }(x),
\label{eq:u1action}
\end{eqnarray}
where the link field $U_{\mu}(x)$ is a complex number with absolute value one and defined on links.
Position $x+ \hat{\mu}$ means the site next to $x$ in the $\mu$ direction.
\begin{eqnarray}
P = \frac{1}{6 N_{\rm site}} \sum_{x, \, \mu > \nu} \mathrm{Re} \left[ U_{\mu}(x) U_{\nu}(x+\hat{\mu})
  U^*_{\mu}(x+\hat{\nu}) U^*_{\nu}(x) \right] 
\end{eqnarray}
is called the plaquette value. $N_{\rm site} =N_s^3 \times N_t$ is the number of sites. 
Then, the gauge action is $S_g = -6 N_{\rm site} \beta P$. 
The gauge field $\theta_{\mu}$ and field strength $ \Theta_{\mu \nu} $ in lattice units are defined as 
\begin{eqnarray}
e^{ i \theta_{\mu} (x) } &=& U_{\mu} (x), \\
e^{ i \Theta_{\mu \nu} (x) } &=& U_{\mu}(x) U_{\nu}(x+\hat{\mu})
         U^*_{\mu}(x+\hat{\nu}) U^*_{\nu}(x)
= e^{ i [ \theta_{\nu} (x+\hat{\mu}) - \theta_{\nu} (x) - \theta_{\mu} (x+\hat{\nu}) + \theta_{\mu} (x) ]}.
\end{eqnarray}
The inverse coupling $\beta$ is given by $\beta=1/g^2$ with the gauge coupling constant $g$.
The gauge field in the continuum theory $A_{\mu}(x)$ is defined as 
\begin{equation}
 \theta_{\mu}(x) = a g A_{\mu}(x) .
\end{equation}
Here, $a$ is the lattice spacing. The field strength in the continuum theory, 
$F_{\mu \nu} = \partial_{\mu} A_{\nu} - \partial_{\nu} A_{\mu}$ 
corresponds to 
\begin{equation}
\Theta_{\mu \nu} = a^{2} g F_{\mu \nu} .
\label{u1fist}
\end{equation}
Because $\Theta_{\mu \nu}$ is small when $a$ is small,  
the path integral can be evaluated around the minimum at $\Theta_{\mu \nu}=0$.
We perform Taylor expansion of Eq.~(\ref{eq:u1action}) around $\Theta_{\mu \nu} \approx 0$, 
\begin{equation}
\cos \Theta_{\mu \nu} \longrightarrow 1- \frac{1}{2} \Theta^{2}_{\mu \nu} .
\label{eq:cosq}
\end{equation}
In the continuum limit, the gauge action becomes
\begin{equation}
 S = \frac{1}{4} \int d^{4} x \, F_{\mu \nu} F_{\mu \nu}  + {\rm const.}
\end{equation}
This is consistent with the action of the continuum theory.

In the continuum limit of the $U(1)$ gauge theory, the flow equation is given by 
\begin{equation}
\frac{\partial B_\mu}{\partial t} (t,x)= \frac{\partial G_{\nu\mu}}{\partial x_\nu} (t,x) = -\frac{\delta S_g}{\delta B_\mu}.
\label{eq:u1flow}
\end{equation}
The flow equation changes the gauge field to minimize the gauge action.
Therefore, we define the flow equation for the gauge field in lattice unit $\theta_\mu$ as 
\begin{equation}
\frac{\partial \theta_\mu^{(t)}(x)}{\partial (t/a^2)}
= -g^2 \frac{\delta S_g}{\delta \theta_\mu^{(t)}} 
= \sum_{\nu=1}^4 \left( \sin \Theta_{\mu \nu}^{(t)}(x) - \sin \Theta_{\mu \nu}^{(t)}(x-\hat{\nu}) \right),
\label{eq:coflow}
\end{equation}
where $t/a^2$ is a dimensionless combination of the flow time, and 
the flowed operators at the flow time are denoted as $\theta_\mu^{(t)}$ and $\Theta_{\mu \nu}^{(t)}$.
We use this flow equation in the following sections except Sec.~\ref{sec:noncompact}.\footnote{
One may use an alternative definition that simply substitutes $a^2 g G_{\nu\mu} = \Theta_{\nu\mu}^{(t)}$ into Eq.~(\ref{eq:u1flow}) to discretize the flow equation.
The difference between such a naive flow equation and Eq.~(\ref{eq:coflow}) is whether the compactness of the gauge group is considered or not.
Without the compactness, the gradient flow doesn't work well.
In Sec.~\ref{sec:noncompact}, we will discuss the importance of this compactness of the flow equation.
}

\subsection*{Numerical simulations}

\begin{figure}[tb]
\begin{minipage}{0.47\hsize}
\begin{center}
\vspace{0mm}
\includegraphics[width=8.1cm]{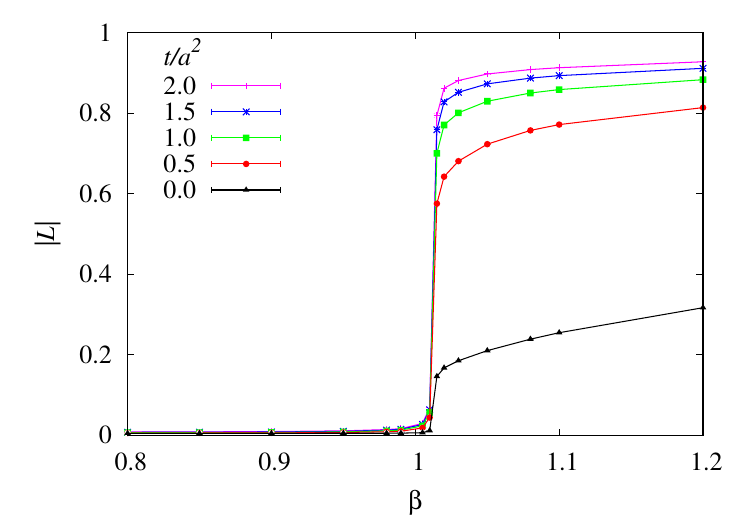}
\vspace{-11mm}
\end{center}
\caption{Absolute value of Polyakov loop as a function of $\beta$ at some flow time $t/a^2$ on a $32^3 \times 8$ lattice.}
\label{fig:plflow-b}
\end{minipage}
\hspace{2mm}
\begin{minipage}{0.47\hsize}
\begin{center}
\vspace{0mm}
\includegraphics[width=8.1cm]{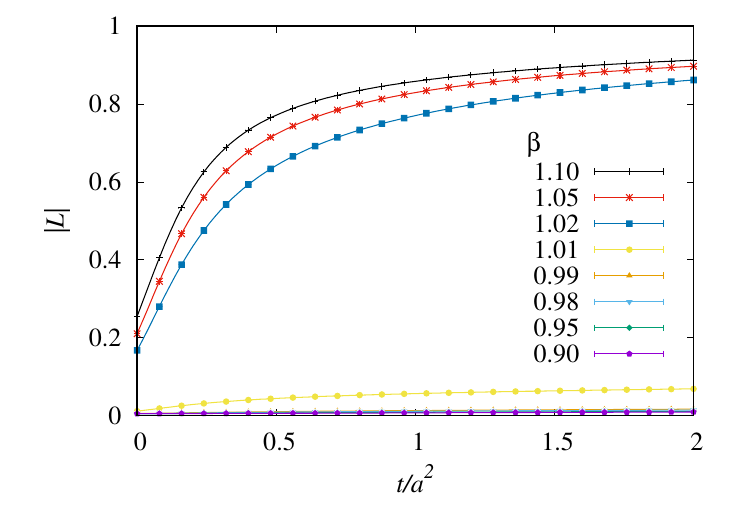}
\vspace{-11mm}
\end{center}
\caption{Absolute value of Polyakov loop as a function of flow time, measured at various $\beta$ on a $32^3 \times 8$ lattice. }
\label{fig:plflow-t}
\end{minipage}
\end{figure}

\begin{figure}[tb]
\begin{minipage}{0.47\hsize}
\begin{center}
\vspace{0mm}
\includegraphics[width=8.1cm]{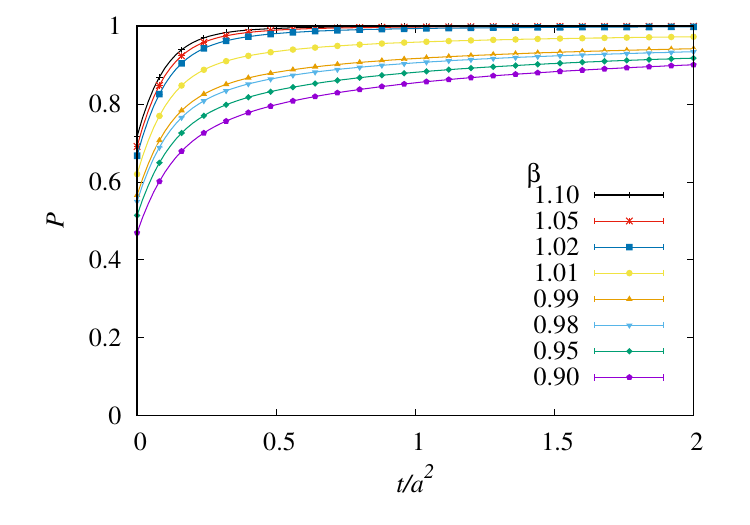}
\vspace{-11mm}
\end{center}
\caption{Plaquette as a function of flow time, computed at various $\beta$ on a $32^3 \times 8$ lattice. }
\label{fig:pqflow}
\end{minipage}
\hspace{2mm}
\begin{minipage}{0.47\hsize}
\begin{center}
\vspace{0mm}
\includegraphics[width=8.1cm]{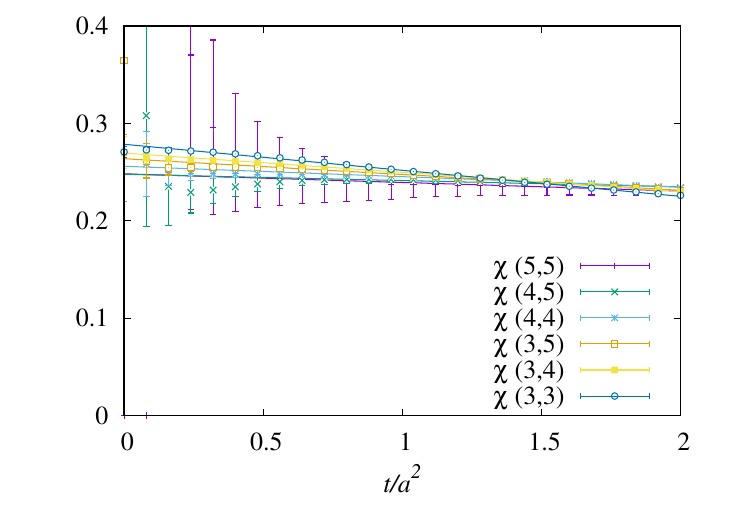}
\vspace{-11mm}
\end{center}
\caption{The flow time dependence of Creutz ratios $\chi (i,j)$ in the confinement phase, measured at $\beta=0.99$ on a $20^4$ lattice.}
\label{fig:creutz-t}
\end{minipage}
\end{figure}

We perform simulations of the $U(1)$ lattice gauge theory to investigate changes in physical quantities during the gradient flow.
The $U(1)$ lattice gauge theory has a deconfinement phase transition. 
The critical $\beta$ is about $\beta_c=1.01$, which depends on the lattice size.
The confinement phase is below $\beta_c$ and the deconfinement phase is above $\beta_c$. 
The expectation value of the Polyakov loop is an order parameter of the confinement phase transition and is defined as 
\begin{eqnarray}
\left\langle L \right\rangle = \left\langle \frac{1}{N_s^3} \sum_{\vec{x}} \exp \left\{ i \sum_{j=0}^{N_{t}-1} \theta_4 (\vec{x} + j \hat4) \right\} \right\rangle .
\end{eqnarray}
From the Polyakov loop, the free energy of a charged particle is given by $-T \ln \langle L \rangle$, where $T$ is the temperature. 

Figures \ref{fig:plflow-b} and \ref{fig:plflow-t} show the flow time dependence of the Polyakov loop.
The expectation value of the absolute value of the Polyakov loop\footnote{
Since the expectation of the Polyakov loop operator itself is always zero due to the $U(1)$ center symmetry, we compute the expectation of the absolute value of the operator.
We will discuss the center symmetry in Sec.~\ref{sec:center}.
} 
is plotted as a function of $\beta$ in Fig.~\ref{fig:plflow-b}.
The black, red, green, blue, and magenta symbols are the results measured at flow time $t/a^2=0.0$, 0.5, 1.0, 1.5, and 2.0, respectively.
The Polyakov loop as a function of flow time $t/a^2$ for each $\beta$ is plotted in Fig.~\ref{fig:plflow-t}.
We generate configurations using a usual pseudo-heat-bath algorithm \cite{Creutz:1983ev}.
Measurements are taken every 1000 updates for each link.
For each configuration, we solve the flow equation of Eq.~(\ref{eq:coflow}) that considers the compactness of the link field $U_{\mu}(x)$.
The lattice size is $N_{\rm site}= 32^3 \times 8$.
The number of independent configurations is 2000 for each $\beta$.
We impose periodic boundary conditions in all directions.
We find from Figs.~\ref{fig:plflow-b} and \ref{fig:plflow-t} that the Polyakov loop increases with the gradient flow in the deconfinement phase; however in the confinement phase, the Polyakov loop remains zero.
This means that the confinement and deconfinement properties do not change by the gradient flow.
We also plot the results of the expectation value of the plaquette in Fig.~\ref{fig:pqflow}.
As the flow time increases, the plaquette approaches one.
Since plaquette is related to the field strength $F_{\mu \nu}$ in the continuum theory by the equation $P \approx 1- \int F_{\mu \nu} F_{\mu \nu} d^4 x \ g^2/(24 V_4)$, we can see that the field strength becomes weaker as the flow time progresses in both confinement and deconfinement phases, where $V_4$ is the space-time volume.

We moreover compute the Creutz ratio $\chi (i,j)$, which is a quantity for simply calculating the string tension,
\begin{eqnarray}
\chi (i,j) = - \ln \left[ \frac{W(i,j) \, W(i+1,j+1)}{W(i+1,j) \, W(i,j+1)} \right].
\label{eq:creutz}
\end{eqnarray}
The string tension is the proportionality constant of the linear potential between static fermions.
If Wilson loops obey the area law, the Creutz ratio is equal to the string tension.
Figure~\ref{fig:creutz-t} shows the flow time dependence of the Creutz ratio at $\beta=0.99$ in confinement phase on a $20^4$ lattice.
The number of configurations is 10000 for each $\beta$.
Since the potential increases linearly at long distances in the confinement phase, larger Wilson loops demonstrate the area law.
Thus, the Creutz ratio $\chi (i,j)$ approaches the string tension as the size $(i,j)$ increases.
This figure shows that the Creutz ratio does not change much with the gradient flow.
We, moreover, find that, as the gradient flow progresses, the size dependence of the Creutz ratio becomes smaller and converges to a certain value.
Also, the Creutz ratio with a larger size has less flow time dependence.

As we expected, the statistical error of the Creutz ratios decreases with the gradient flow.
It may be possible to calculate the string tension by extrapolating to $t/a^2 =0$ using Creutz ratio data at finite $t/a^2$ with large $(i,j)$ that cannot be calculated before the gradient flow due to large statistical errors.
Since $\chi(i,j)$ with small statistical error varies linearly with respect to $t/a^2$, we fit $\chi(i,j)$ with a straight line for each $(i,j)$.
In Fig.~\ref{fig:creutz-t}, the solid lines are the results of the fit function.
We will discuss this issue in more detail in Sec.~\ref{sec:string}. 
These features suggest the usefulness of the gradient flow for numerical computation of lattice gauge theories.
However, it is curious that the string tension hardly changes even though the field strength weakens with the gradient flow.

\section{Monopole in $U(1)$ lattice gauge theory}
\label{sec:monopole}

The existence of topological quantities such as magnetic monopoles is expected to be behind the fact that the confinement property is maintained even with coarse graining by the gradient flow.
In the compact $U(1)$ lattice gauge theory, magnetic monopoles can be defined in the following way \cite{DeGrand:1980eq}, which corresponds to the magnetic monopoles caused by the compactness of the gauge group \cite{Banks:1977cc}.
Hereafter, for the differentiation of variables defined in lattice units, such as $\theta_{\mu}$, $\Theta_{\mu \nu}$, $k_{\mu}$, $n_{\mu \nu}$, etc., the forward derivative $\partial_{\mu} f(x)$ 
and the backward derivative $\partial'_{\mu} f(x)$ in a lattice unit are defined as 
\begin{eqnarray}
\partial_{\mu} f(x) = f(x+ \hat{\mu}) - f(x), \hspace{5mm} 
\partial'_{\mu} f(x) = f(x) - f(x- \hat{\mu}) ,
\end{eqnarray}
for any function $f (x)$, where $x+ \hat{\mu}$ means the site next to $x$ in the $\mu$ direction.

Since 
$\Theta_{\mu \nu} = \partial_{\mu} \theta_{\nu} - \partial_{\nu} \theta_{\mu}$, 
the range of $\theta_{\mu}$ is defined as $- \pi < \theta_{\mu} \leq \pi$, and then the range of $\Theta_{\mu \nu}$ is $-4 \pi < \Theta_{\mu \nu} \leq 4 \pi$.
In this case, $- \cos \Theta_{\mu \nu}$ in $S_g$ has minimum value at $\Theta_{\mu \nu} =0,$ $\pm 2 \pi$, and $\pm 4 \pi$, and the approximation of Eq.~(\ref{eq:cosq}) does not hold except near $\Theta_{\mu \nu} =0$.
To avoid this problem, we define the quantity $\bar{\Theta}_{\mu \nu}$ as follows, and 
it is more appropriate to regard $\bar{\Theta}_{\mu \nu}$ as $F_{\mu \nu}$ in the continuum theory rather than $\Theta_{\mu \nu}$:
\begin{equation}
\Theta_{\mu \nu} = \partial_{\mu} \theta_{\nu} 
- \partial_{\nu} \theta_{\mu} = \bar{\Theta}_{\mu \nu} + 2 \pi n_{\mu \nu} ,
\label{eq:thbar}
\end{equation}
with
\begin{equation}
 \bar{\Theta}_{\mu \nu} = a^{2} g F_{\mu \nu} , 
\end{equation}
where $n_{\mu \nu}$ is an integer and $- \pi < \bar{\Theta}_{\mu \nu} \leq \pi$.
Moreover, when converting from $U_{\mu}$ to $\theta_{\mu}$, the ambiguity of integer multiples of $2\pi$ does not affect $\bar{\Theta}_{\mu \nu}$.

When the field strength is defined in this way, $\bar{\Theta}_{\mu \nu}$ does not satisfy the Bianchi identity.
From the definition, 
\begin{eqnarray}
 \epsilon_{\mu \nu \rho \sigma} \partial_{\nu} \Theta_{\rho \sigma}
 = \epsilon_{\mu \nu \rho \sigma} \partial_{\nu} \bar{\Theta}_{\rho \sigma}
 + 2 \pi \epsilon_{\mu \nu \rho \sigma} \partial_{\nu} n_{\rho \sigma} = 0 .
\end{eqnarray}
Thus, 
\begin{equation}
 \epsilon_{\mu \nu \rho \sigma} \partial_{\nu} \bar{\Theta}_{\rho \sigma}
 = -2 \pi \epsilon_{\mu \nu \rho \sigma} \partial_{\nu} n_{\rho \sigma}.
\label{eq:bianchi}
\end{equation}
The right-hand side of Eq.~(\ref{eq:bianchi}) is not always zero.
We then define the magnetic monopole current as follows \cite{DeGrand:1980eq}: 
\begin{equation}
k_{\mu} (x) = \frac{1}{4 \pi} 
\epsilon_{\mu \nu \rho \sigma} \partial_{\nu} \bar{\Theta}_{\rho \sigma} (x)
= - \frac{1}{2} \epsilon_{\mu \nu \rho \sigma} \partial_{\nu} n_{\rho \sigma} (x).
\label{eq:monopole}
\end{equation}
Here, we note that $k_{\mu} (x)$ is an integer, which corresponds to Dirac's quantization condition.
Furthermore, $k_{\mu} (x)$ satisfies the continuity equation,\footnote{
If we want to define the continuity equation using the backward derivative on a lattice, we need to shift the definition of the monopole current by one site, i.e., 
if $k_{\mu} (x) \to k'_{\mu} (x) = k_{\mu} (x+\hat{\mu})$, then $\partial'_{\mu} k'_{\mu} (x) =0.$
The current $k'_{\mu}$ is conserved at $x$ on the lattice. 
Moreover, it is more appropriate to express Eq.~(\ref{eq:mon-ds}) by the backward derivative.
if $^{\ast} \hspace{-1mm} n_{\mu \nu} (x) \to \, 
^{\ast} \hspace{-1mm} n'_{\mu \nu} (x) = \, ^{\ast} \hspace{-1mm} n_{\mu \nu} (x+\hat{\mu}+\hat{\nu})$, 
then $k'_{\mu} (x) = \partial'_{\nu} \, ^{\ast} \hspace{-1mm} n'_{\nu \mu} (x).$
This means that $k'_{\mu}$ is the boundary of $^{\ast} \hspace{-1mm} n'_{\nu \mu}$.
\label{footnote3}
} 
\begin{equation}
\partial_{\mu} k_{\mu} =0.
\label{eq:contk}
\end{equation}
Thus, the monopole current forms a closed loop.
When the monopole current is integrated over the entire space-time under the periodic boundary condition, the integrated value becomes zero.
The dual of $n_{\mu \nu}(x)$ is defined by 
$ ^{\ast} \hspace{-1mm} n_{\mu \nu} (x) = 
\frac{1}{2} \epsilon_{\mu \nu \rho \sigma} n_{\rho \sigma}(x)$.
It satisfies the equation, 
\begin{eqnarray}
k_{\mu}(x) = \partial_{\nu} \hspace{1mm} ^{\ast} 
\hspace{-1mm} n_{\nu \mu} (x).
\label{eq:mon-ds}
\end{eqnarray}
Because the boundary of $^{\ast} \hspace{-1mm} n_{\mu \nu} (x)$ is the monopole current, we call $^{\ast} \hspace{-1mm} n_{\mu \nu} (x)$ Dirac string.
This Dirac string corresponds to an infinitely long and thin solenoid connected to Dirac's magnetic monopole.

When we perform a global change of link field, $U_{\mu}(x) \to e^{\alpha_{\mu}} U_{\mu}(x) = e^{i \alpha_{\mu}} e^{i \theta_{\mu} (x)}$ for any real number $\alpha_{\mu}$, the gauge field $\theta_4$ does not become $\theta_{\mu} + \alpha_{\mu}$ because the range is $-\pi <\theta_{\mu} \leq \pi$, but becomes $\theta_{\mu} + \alpha_{\mu} + 2 \pi n$ with an extra integer multiple of $2 \pi$ added.
Then, in the definition: 
$\partial_{\mu} \theta_{\nu} - \partial_{\nu} \theta_{\mu} 
= \bar{\Theta}_{\mu \nu} + 2 \pi n_{\mu \nu}$, 
$\Theta_{\mu \nu} (x)$ can change by an integer multiple of $2 \pi$, but $\bar{\Theta}_{\mu \nu} (x)$ 
does not change.
Since $k_{\mu} (x)$ is defined by $\bar{\Theta}_{\mu \nu} (x)$, the monopole current is invariant under this transformation.
On the other hand, this transformation changes the Dirac string $^{\ast} \hspace{-1mm} n_{\mu \nu} (x)$. 
Similarly for the gauge transformation, $\bar{\Theta}_{\mu \nu} (x)$ and $k_{\mu} (x)$ are invariant and $^{\ast} \hspace{-1mm} n_{\mu \nu} (x)$ changes.
Therefore, the Dirac string is not a physical quantity.

The magnetic monopole is expected to exist stably against coarse graining.
For the case of the gradient flow considering compactness of $U_{\mu}(x)$,
$- \cos \Theta_{\mu \nu}$ becomes small when we solve the flow equation Eq.~(\ref{eq:coflow}).
Then, $\bar{\Theta}_{\mu \nu}$ will be small, but the integer variable $n_{\mu \nu}$ will not change much in Eq.~(\ref{eq:thbar}).
Therefore, the monopole is expected not to disappear even if the gauge field is coarse grained by the gradient flow in the confinement phase.

\subsection*{Numerical simulation}

\begin{figure}[tb]
\begin{center}
\vspace{0mm}
\includegraphics[width=8.1cm]{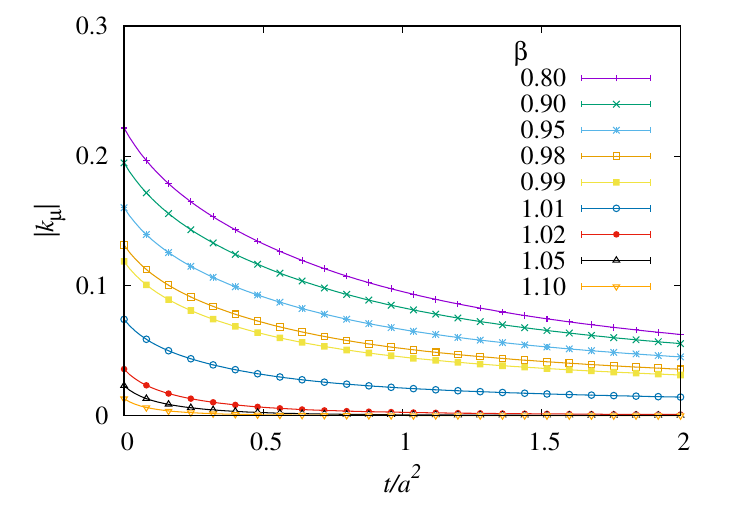}
\hspace{0mm}
\includegraphics[width=8.1cm]{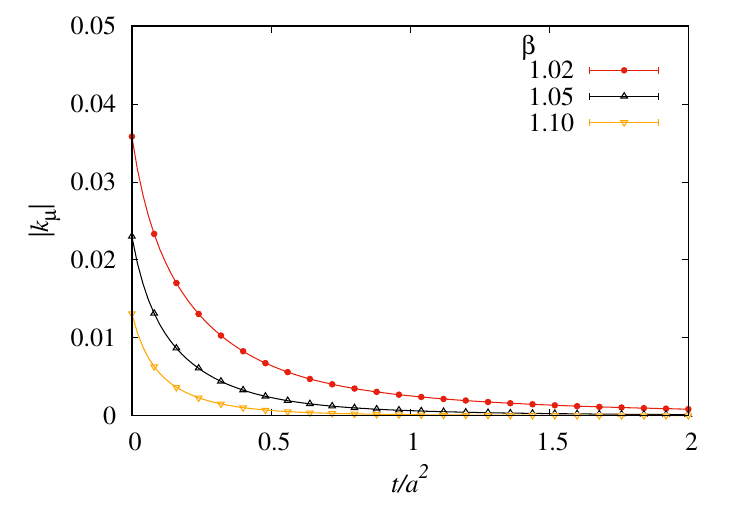}
\vspace{-11mm}
\end{center}
\caption{Density of monopoles $(4N_{\rm site})^{-1} \sum_{x, \mu} |k_{\mu}(x)|$ as a function of flow time for various $\beta$ on a $32^3 \times 8$ lattice.
The results in the deconfinement phase only are plotted in the right panel. 
}
\label{fig:monflow}
\end{figure}

\begin{figure}[tb]
\begin{minipage}{0.47\hsize}
\begin{center}
\vspace{0mm}
\includegraphics[width=8.1cm]{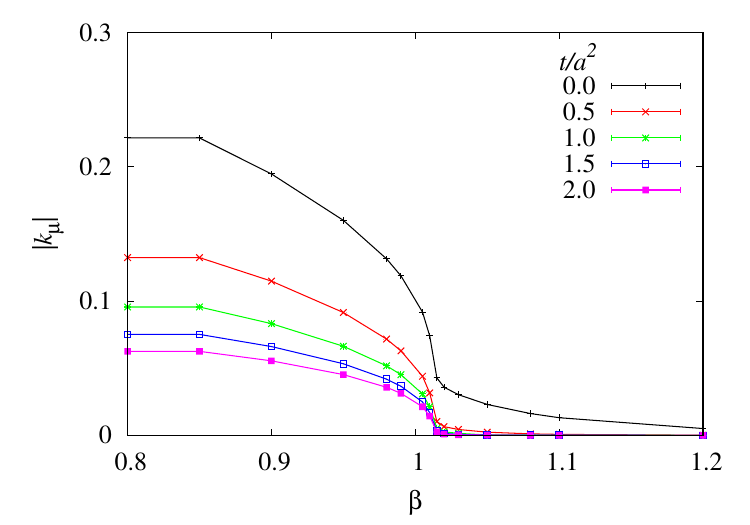}
\vspace{-11mm}
\end{center}
\caption{Density of monopoles as a function of $\beta$ on a $32^3 \times 8$ lattice. Black, red, green, blue and magenta lines are the results of $t/a^2=0.0$, 0.5, 1.0, 1.5, and 2.0, respectively.}
\label{fig:monflow-b}
\end{minipage}
\hspace{2mm}
\begin{minipage}{0.47\hsize}
\begin{center}
\vspace{0mm}
\includegraphics[width=8.1cm]{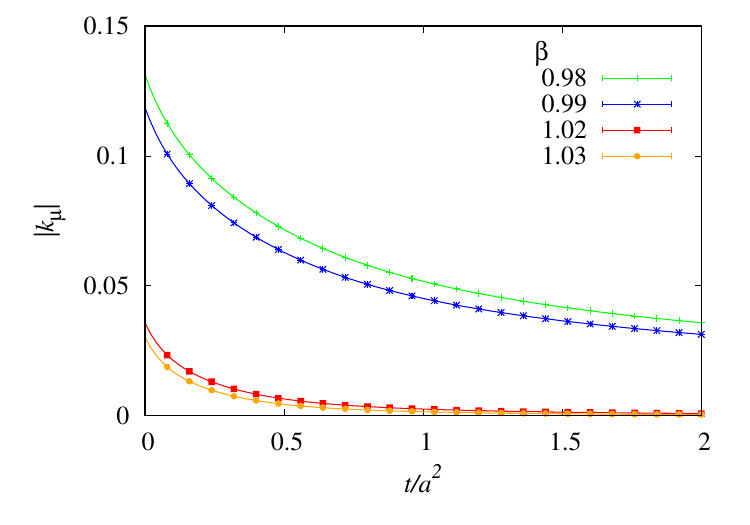}
\vspace{-11mm}
\end{center}
\caption{Density of monopoles as a function of $t/a^2$ at $\beta=0.98$, 0.99, 1.02,  and 1.03 on a $20^4$ lattice.}
\label{fig:monflow20}
\end{minipage}
\end{figure}

By performing Monte Carlo simulations, we investigate how the magnetic monopole changes with the gradient flow.
The monopoles are computed on the configurations used in the calculations in Sec.~\ref{sec:gradient}.
Because the space-time average of the monopole current is zero due to Eq.~(\ref{eq:contk}), we focus on the space-time average of the absolute value of  $k_{\mu} (x)$.
The results of the expectation values of the density $(4N_{\rm site})^{-1} \sum_{x, \mu} |k_{\mu}(x)|$ on the $32^3 \times 8$ lattice are plotted in the left panel of Fig.~\ref{fig:monflow}.
As seen from this figure, the number of monopoles is large at $\beta \le 1.01$ in the confinement phase before the flow. Then, the gradient flow causes the number of monopoles to slowly decrease, but it seems to never reach zero.
On the other hand, there are only a few monopoles at $\beta \ge 1.02$ in the deconfinement phase before the gradient flow, and monopoles disappear immediately after gradient flow.
In the right panel of Fig.~\ref{fig:monflow}, the range of the vertical axis is narrowed to show only the results for the deconfinement phase.
We also plot the number of monopoles as a function of $\beta$ for each flow time in Fig.~\ref{fig:monflow-b}.
The number of monopoles at large flow times changes dramatically at the critical $\beta$ from a finite value in the confinement phase to zero in the deconfinement phase.
Figure \ref{fig:monflow20} is the result on a symmetric lattice with $N_{\rm site} =20^4$ at $\beta=0.98$ (green), 0.99 (blue), 1.02 (red), and 1.03 (yellow). 
The configuration at $\beta=0.99$ is used in the study of the string tension.
The number of configurations at the other $\beta$ is 4000.
This is qualitatively the same result as the $32^3 \times 8$ lattice.
These results suggest that there is a relationship between the monopole not vanishing in the confinement phase and the preservation of the confinement properties.

\section{Gradient flow with noncompact flow equation}
\label{sec:noncompact}

So far, we have used Eq.~(\ref{eq:coflow}) as the flow equation, but alternative flow equations are also possible.
By simply discretizing Eq.~(\ref{eq:u1flow}) using the relation of $a^2 g G_{\mu\nu} = \Theta_{\mu\nu}^{(t)}$, we can also define the flow equation as
\begin{equation}
\frac{\partial \theta_\mu^{(t)}(x)}{\partial (t/a^2)}
 = \sum_{\nu}  \left( \Theta_{\mu \nu}^{(t)}(x) - \Theta_{\mu \nu}^{(t)}(x-\hat{\nu}) \right) ,
\label{eq:noncoflow}
\end{equation}
with $\Theta_{\mu \nu} = \partial_{\mu} \theta_{\nu} - \partial_{\nu} \theta_{\mu}$.
The difference between Eq.~(\ref{eq:coflow}) and Eq. (\ref{eq:noncoflow}) is the consideration of the periodicity of $\cos \Theta_{\mu \nu}$ in the action.
The right-hand side of Eq.~(\ref{eq:noncoflow}) is equal to the functional derivative of $S_{\rm NC}$, which is the action of the noncompact $U(1)$ gauge theory.
\begin{equation}
\frac{\partial \theta_{\mu}^{(t)}(x)}{\partial (t/a^2)}
 = -g^2 \frac{\delta S_{\rm NC}}{\delta \theta_{\mu}^{(t)}} , 
\hspace{5mm} {\rm with} \hspace{5mm}
S_{\rm NC} = \beta \sum_{x, \mu > \nu} \frac{1}{2} \Theta_{\mu \nu}^2(x).
\end{equation}
Thus, this flow equation changes the gauge field to minimize $S_{\rm NC}$.
Unlike the compact action Eq.~(\ref{eq:u1action}), this $S_{\rm NC}$ has no periodicity in $\Theta_{\mu \nu}$.

\begin{figure}[tb]
\begin{minipage}{0.47\hsize}
\begin{center}
\vspace{0mm}
\includegraphics[width=8.1cm]{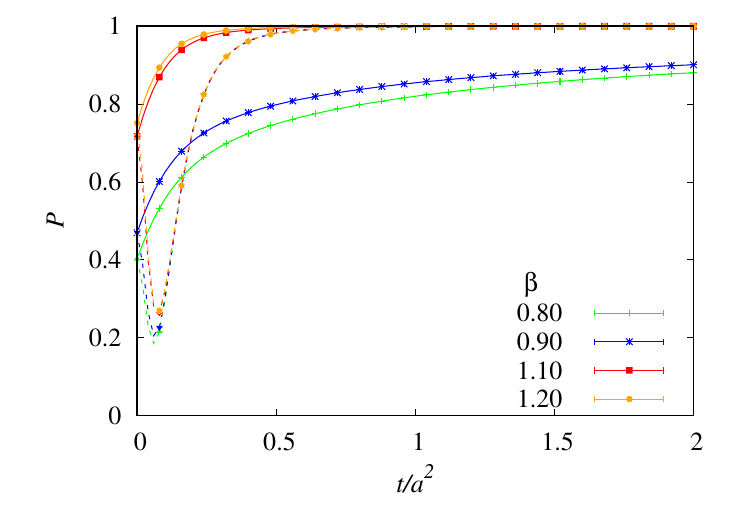}
\vspace{-11mm}
\end{center}
\caption{Flow time dependence of plaquette using the noncompact (dashed line) and the compact (solid line) flow equations, measured at $\beta=0.80$ (green), 0.90 (blue), 1.10 (red) and 1.20 (yellow) on a $16^4$ lattice. }
\label{fig:nonpqflow}
\end{minipage}
\hspace{2mm}
\begin{minipage}{0.47\hsize}
\begin{center}
\vspace{0mm}
\includegraphics[width=8.1cm]{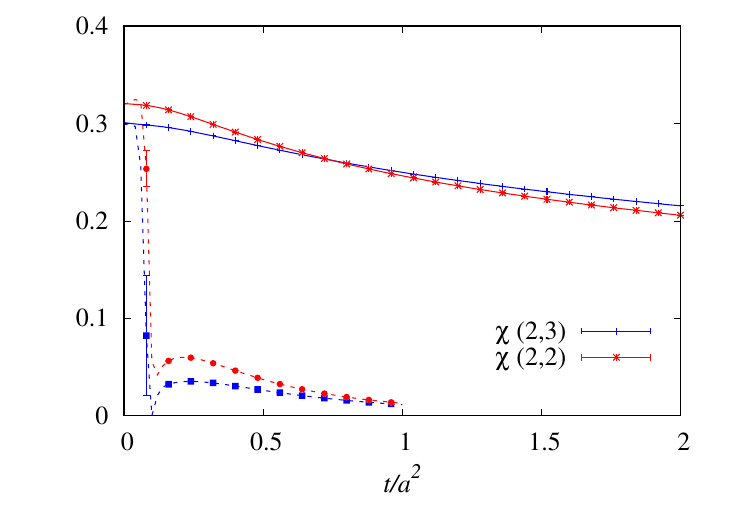}
\vspace{-11mm}
\end{center}
\caption{Creutz ratios $\chi(2, 2)$ (red) and $\chi(2,3)$ (blue) as functions of flow time at $\beta=0.99$ on a $16^4$ lattice using the noncompact (dashed line) and the compact (solid line) flow equations.}
\label{fig:nonstrflow}
\end{minipage}
\end{figure}

\begin{figure}[tb]
\begin{minipage}{0.47\hsize}
\begin{center}
\vspace{0mm}
\includegraphics[width=8.1cm]{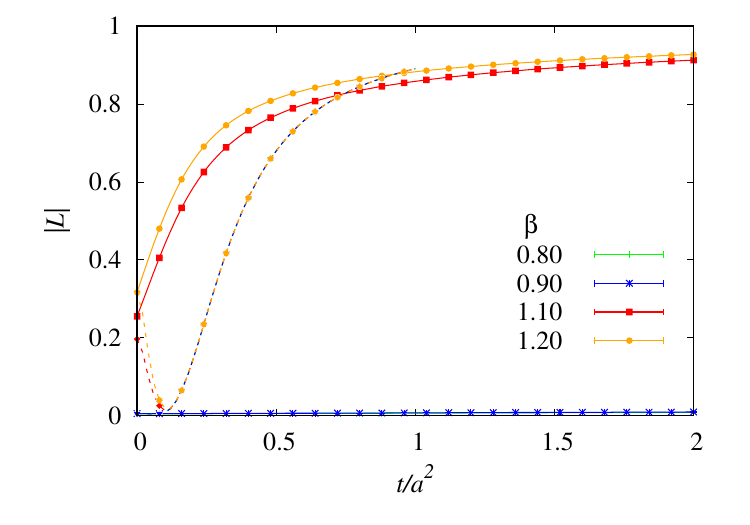}
\vspace{-11mm}
\end{center}
\caption{Absolute values of Polyakov loop as functions of flow time at $\beta=0.80$, 0.90, 1.10 and 1.20 on a $32^3 \times 8$ lattice, using the noncompact (dashed line) and the compact (solid line) flow equations.}
\label{fig:nonplflow}
\end{minipage}
\hspace{2mm}
\begin{minipage}{0.47\hsize}
\begin{center}
\vspace{0mm}
\includegraphics[width=8.1cm]{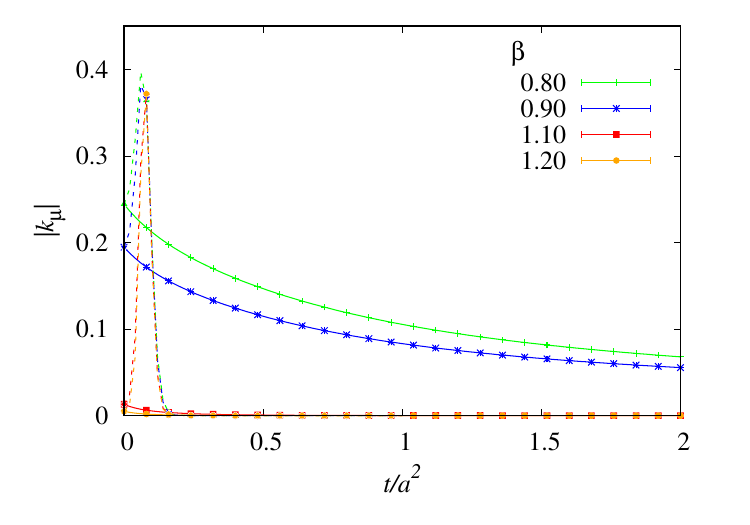}
\vspace{-11mm}
\end{center}
\caption{Density of monopoles as a function of flow time at $\beta=0.80$, 0.90, 1.10, and 1.20 on a $16^4$ lattice. 
The dashed and solid lines are the results by the noncompact and compact flow equations, respectively.}
\label{fig:nonmonflow}
\end{minipage}
\end{figure}

We investigate how the confinement properties and the magnetic monopole change with gradient flow for two cases of the gradient flow defined by the compact flow equation Eq.~(\ref{eq:coflow}) and the noncompact flow equation Eq.~(\ref{eq:noncoflow}) performing Monte Carlo simulations.
The results of these two types of flow equations are drastically different.

We plot the results of plaquette values measured on a $16^4$ lattice in Fig.~\ref{fig:nonpqflow}. 
Configurations are generated by the action of the compact $U(1)$ lattice gauge theory, Eq.~(\ref{eq:u1action}).
The number of configurations is 10000 for each $\beta$.
The solid lines are the case using the compact flow equation, and the dashed lines are the noncompact case.
These two results are qualitatively very different.
As discussed in Sec.~\ref{sec:gradient}, for the compact flow, the plaquette values monotonically approach one as the flow progresses.
However, in the case of the noncompact flow, the plaquette values decrease once immediately after the flow starts, the difference between the confinement phase ($\beta=0.80$, 0.90) and the deconfinement phase ($\beta=1.10$, 1.20) disappears, and they approach one as the flow time increases.

A more pronounced difference in the Creutz ratio results is seen when two different flow equations are used.
In Fig.~\ref{fig:nonstrflow}, we plot the results of the Creutz ratios $\chi(2, 2)$ (red) and $\chi(2,3)$ (blue) measured at $\beta=0.99$ on a $16^4$ lattice.
The solid lines and the dashed lines are for the compact and the noncompact cases, respectively.
It can be seen that the Creutz ratio vanishes immediately when the gradient flow is performed using the noncompact flow equation.
This indicates that the confinement property is lost by using the noncompact flow equation.

We moreover plot the absolute value of Polyakov loop as a function of the flow time in Fig.~\ref{fig:nonplflow}.
The solid lines are the results by the compact flow equation, which have been shown in Sec.~\ref{sec:gradient}. 
As discussed in Sec.~\ref{sec:gradient}, the qualitative properties of zero or finite values do not change with the gradient flow.
The dashed lines are the results by the noncompact flow equation Eq.~(\ref{eq:noncoflow}), measured on a $32^3 \times 8$ lattice.
This figure shows that as the noncompact flow progresses, the Polyakov loop decreases until $t/a^2 \approx 0.1$, where the $\beta$ dependence disappears.
After that, the Polyakov loop increases with the same value for all $\beta$ (i.e. $\beta$ in both phases) as the flow time increases.
This also indicates that the nature of confinement is lost by the noncompact gradient flow.

The magnetic monopoles discussed in the previous section appear due to the compactness of $U(1)$ group elements $U_{\mu}(x)$.
Magnetic monopoles are expected to be preserved when coarse grained using the compact flow equation.
However, the noncompact flow equation Eq.~(\ref{eq:noncoflow}) breaks the compactness of $U_{\mu}(x)$, unlike the compact flow equation Eq.~(\ref{eq:coflow}).
There is no reason why magnetic monopoles should not vanish if coarse grained using a flow equation without the compactness.
We compute the expectation values of $(4N_{\rm site})^{-1} \sum_{x, \mu} |k_{\mu}(x)|$, and plot the results of the $16^4$ lattice in Fig.~\ref{fig:nonmonflow}.
The solid lines in Fig.~\ref{fig:nonmonflow} are the density of monopoles for the case of  Eq.~(\ref{eq:coflow}) considering the compactness.
As discussed in Sec.~\ref{sec:monopole}, the monopole does not disappear even with gradient flow in the confinement phase.
However, the dashed lines in Fig.~\ref{fig:nonmonflow} are the results for the case of noncompact gradient flow Eq.~(\ref{eq:noncoflow}).
The number of monopoles increases once after the flow starts, and the number of monopoles becomes the same for all $\beta$ at $t/a^2 \approx 0.1$.
After that, the number of monopoles immediately decreases to zero for all $\beta$.
The results obtained in this section indicate that it is important to use the flow equation considering the compactness of the gauge group so that the confinement properties are not lost under the gradient flow.
These results also suggest that the disappearance of the monopoles corresponds to the disappearance of the confinement properties.

\section{String tension}
\label{sec:string}

The most important property of confinement is the linear potential between particles.
The effective potential $V(r)$ at the distance $r$ between particles can be calculated from Wilson loops $W(r, t)$ by the relation $W(r, t) = \exp[ -t V(r) ]$ when $t$ is large.
If the effective potential has a linear term $V(r) \sim \sigma r$ at long distances, where $\sigma$ is the string tension, Wilson loops show the area law, $W(r, t) \sim \sigma r t$, for large $r$ and $t$.
The area law can be derived in the strong coupling limit of lattice gauge theories. 
In the proof of the area law, the compactness of the gauge group is very important.
Moreover, the area law for Wilson loops is given by monopole condensation in the compact $U(1)$ lattice gauge theory with the Villain approximation \cite{Banks:1977cc}.
So far, we have argued that the magnetic monopoles may also play an important role in the gradient flow.
In this section, we study the flow time dependence of the Wilson loops focusing on the monopole condensation.

The Wilson loop can be separated into contributions from field strength and magnetic monopoles \cite{Stack:1991zp}.
We introduce an external current $J_{\mu}(x)$ that takes $\pm 1$ along the Wilson loop and an antisymmetric tensor $M_{\mu \nu}(x)$ that takes $\pm 1$ in the area surrounded by the Wilson loop and satisfies $J_{\nu}(x) = \partial'_{\mu} M_{\mu\nu}(x)$.
The Wilson loop is given as 
\begin{eqnarray}
W = \exp \left\{i \sum_{x} \theta_{\mu}(x) J_{\mu}(x) \right\} 
      = \exp \left\{ - \frac{i}{2} \sum_{x} \Theta_{\mu \nu}(x) M_{\mu \nu}(x) \right\} . 
\end{eqnarray}
Using the identity, 
\begin{eqnarray}
M_{\mu \nu} (x) = - \sum_{x'} D(x-x') \left[ \partial'_{\alpha} 
(\partial_{\mu} M_{\alpha \nu}(x') - \partial_{\nu} M_{\alpha \mu}(x') ) 
+ \frac{1}{2} \epsilon_{\alpha \beta \mu \nu} 
\epsilon_{\lambda \beta \rho \sigma} \partial'_{\alpha} \partial_{\lambda} 
M_{\rho \sigma}(x') \right] .
\end{eqnarray}
the Wilson loop can be decomposed as follows: 
\begin{eqnarray}
W & = & W_f \cdot W_m \label{w12} , \\
W_f & = & \exp \left\{ -i \sum_{x,x'} \partial'_{\mu}\bar{\Theta}_{\mu\nu}(x)
D(x-x')J_{\nu}(x') \right\} , \\
W_m & = & \exp \left\{ 2\pi i \sum_{x,x'} k_{\beta}(x)D(x-x')\frac{1}{2}
\epsilon_{\alpha\beta\rho\sigma}\partial_{\alpha}M_{\rho\sigma}(x') \right\} , 
\label{wmon}
\end{eqnarray}
where $D(x-x')$ is a four-dimensional Coulomb propagator on a lattice, which satisfies 
\begin{eqnarray}
\partial'_{\mu} \partial_{\mu} D(x-x') = - \delta_{x,x'}. 
\label{eq:cprop}
\end{eqnarray}
Then, $\langle W \rangle$ is approximately the product of $\langle W_f \rangle$ and $\langle W_m \rangle$.
$\bar{\Theta}_{\mu \nu}$ corresponds to $F_{\mu \nu}$ in the continuum theory. 
It is known that the string tension calculated only by $W_m$ without $W_f$ almost reproduces the original string tension by numerical calculation \cite{Stack:1991zp}.

The contribution from monopoles to the Wilson loop is equivalent to the Wilson loop with the contribution from the field strength $F_{\mu \nu}$ removed by hand, and is very similar to the Wilson loop in which $F_{\mu \nu}$ is reduced by gradient flow.
We investigate how the contributions from monopoles and $F_{\mu \nu}$ change when the field is coarse grained by the gradient flow, and compare them with the original string tension.

\subsection*{Numerical simulation}

\begin{figure}[tb]
\begin{center}
\vspace{0mm}
\includegraphics[width=8.1cm]{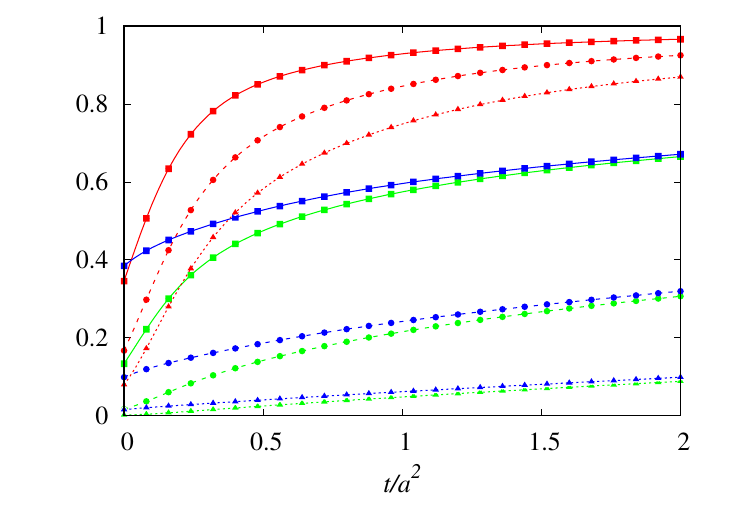}
\hspace{0mm}
\includegraphics[width=8.1cm]{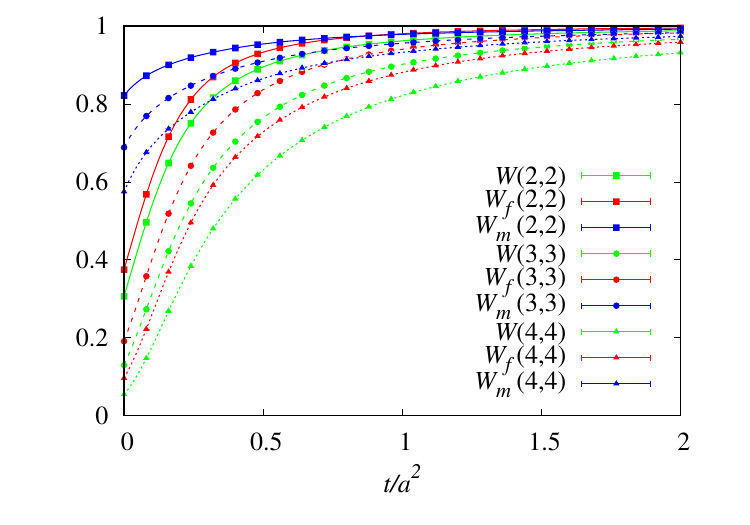}
\vspace{-11mm}
\end{center}
\caption{Flow time dependence of Wilson loops (green), their field strength contributions (red), and their monopole contributions (blue) on a $20^4$ lattice.
The left and right panels are the results at $\beta=0.99$ and $1.02$.
The square, circle, and triangle symbols are $W(2,2)$, $W(3,3)$, and $W(4,4)$, respectively.}
\label{fig:wlflow}
\end{figure}

\begin{figure}[tb]
\begin{minipage}{0.47\hsize}
\begin{center}
\vspace{0mm}
\includegraphics[width=8.1cm]{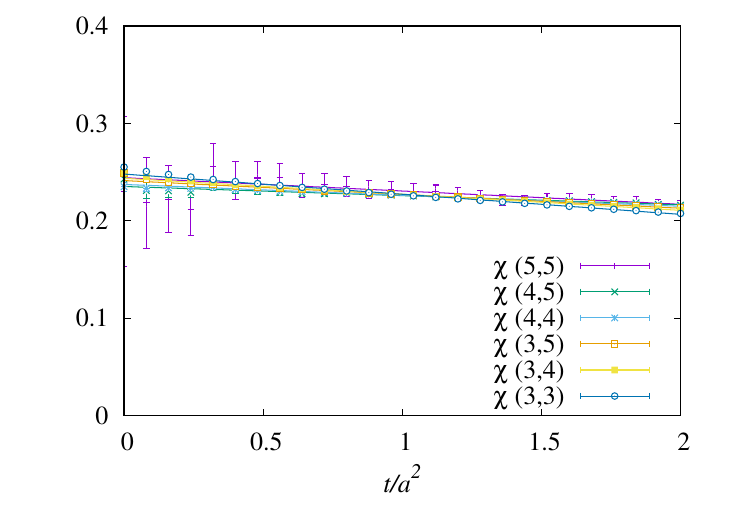}
\vspace{-11mm}
\end{center}
\caption{The monopole contribution of Creutz ratios as functions of flow time at $\beta=0.99$ on a $20^4$ lattice.}
\label{fig:cremonflow}
\end{minipage}
\hspace{2mm}
\begin{minipage}{0.47\hsize}
\begin{center}
\vspace{0mm}
\includegraphics[width=8.1cm]{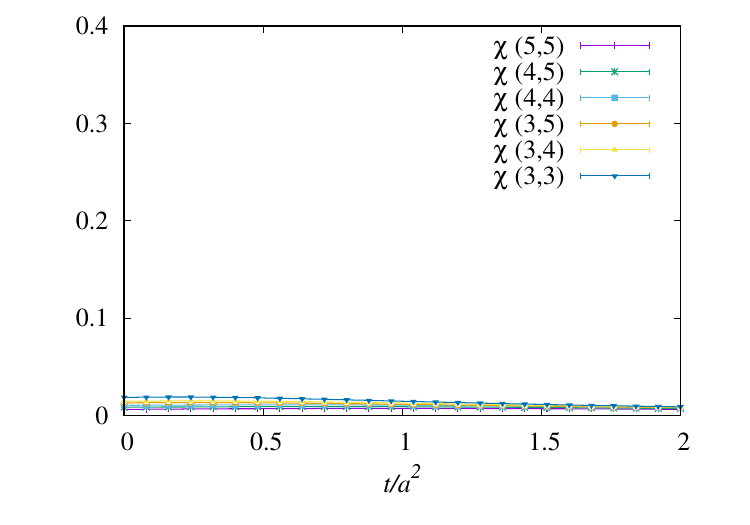}
\vspace{-11mm}
\end{center}
\caption{The $F_{\mu \nu}$ contribution of Creutz ratios as functions of flow time at $\beta=0.99$ on a $20^4$ lattice. }
\label{fig:crephflow}
\end{minipage}
\end{figure}

We decompose the Wilson loop into two contributions and investigate the flow time dependence.
The simulations are performed on a lattice of size $20^4$.
The number of configurations is 10000 for $\beta=0.99$, and 4000 for $\beta=1.02$.
During the gradient flow, we solve the flow equation Eq.~(\ref{eq:coflow}).
In Fig.~\ref{fig:wlflow}, we plot the results of Wilson loops $\langle W(2,2) \rangle, \langle W(3,3) \rangle$ and $\langle W(4,4) \rangle$ as functions of the flow time $t/a^2$ with square, circle, and triangle symbols.
The left and right panels are the results at $\beta=0.99$ (confinement) and $\beta=1.02$ (deconfinement). 
The green, red, and blue symbols are the original Wilson loop, the $F_{\mu \nu}$ contribution, and the monopole contribution, respectively.
These Wilson loops increase monotonically as the flow time increases in Fig.~\ref{fig:wlflow}.

If Wilson loops show the area law, the Creutz ratio $\chi (i,j)$, Eq.~(\ref{eq:creutz}), is equal to the string tension.
From the Wilson loops, we compute the Creutz ratio.
%\begin{eqnarray}
%\chi (i,j) = - \ln \left[ \frac{W(i,j) \, W(i+1,j+1)}{W(i+1,j) \, W(i,j+1)} \right].
%\end{eqnarray}
As discussed in Sec.~\ref{sec:gradient}, Fig.~\ref{fig:creutz-t} shows the flow time dependence of the original Creutz ratio.
In the confinement phase, the Creutz ratio approaches the string tension as the size $(i,j)$ increases, and the size dependence becomes smaller as the gradient flow progresses.
This means that when coarse grained by the gradient flow, the interparticle potential becomes a linear rising potential even at short distances.
Figure~\ref{fig:cremonflow} demonstrates the results obtained by the Wilson loops from monopoles $\langle W_m (i,j) \rangle$.
The statistical error of the Creutz ratio from monopoles is smaller than that of the original Creutz ratio.
These Creutz ratios are independent of the size of Wilson loops within the errors. 
This indicates that both before and after flow, the interparticle potential is a linear function even at short distances.
Moreover, their flow time dependence is smaller than the original one.

\begin{table}[t]
\begin{center}
\caption{Fitting parameters of the Creutz ratios on the $20^4$ lattice.}
\label{tab:chi}
%\small
\begin{tabular}{ccccc}
\hline
\hspace{10mm}  & \multicolumn{2}{c}{original} & \multicolumn{2}{c}{monopole} \\
\hspace{10mm} & \hspace{5mm} $c(i,j)$ \hspace{5mm} & \hspace{5mm} $d(i,j)$ \hspace{5mm} &
\hspace{5mm} $c(i,j)$ \hspace{5mm} & \hspace{5mm} $d(i,j)$ \hspace{5mm} \\
\hline
$\chi (3,3)$ & 0.2787(4)  & -0.0267(2)  & 0.2481(2)  & -0.0207(1)  \\
$\chi (3,4)$ & 0.2696(7)  & -0.0198(4)  & 0.2437(5)  & -0.0164(2)  \\
$\chi (3,5)$ & 0.2639(15) & -0.0162(7)  & 0.2414(9)  & -0.0142(4)  \\
$\chi (4,4)$ & 0.2561(25) & -0.0107(12) & 0.2374(14) & -0.0112(7)  \\
$\chi (4,5)$ & 0.2479(64) & -0.0065(32) & 0.2353(34) & -0.0094(17) \\
$\chi (5,5)$ & 0.248(28)  & -0.009(14)  & 0.244(14)  & -0.014(7)   \\
\hline
\end{tabular}
\end{center}
\end{table}

Since the numerical result of the Creutz ratio with small statistical error seems to be well approximated by a linear function, we fit these Creutz ratios at $\beta=0.99$ with a linear function,
\begin{eqnarray}
\chi(i,j) = c(i,j) + d(i,j) \, t/a^2 ,
\end{eqnarray}
for each size $(i,j)$, where $c(i,j)$ and $d(i,j)$ are the fitting parameters.
The straight lines in Figs.~\ref{fig:creutz-t} and \ref{fig:cremonflow} are the fitting functions.
The results of the fitting parameters are summarized in Table~\ref{tab:chi}.
The second and third columns are the results of the original Creutz ratio, and the fourth and fifth columns are the results of the Creutz ratio from monopoles.
In Appendix \ref{sec:volume}, we discuss the finite volume effect in the Creutz ratios comparing the results on $16^4$ and $20^4$ lattices.
A finite volume effect is visible in Creutz ratios containing Wilson loops with a side length greater than $N_s/3$ for a $N_s^4$ lattice.
Thus, the table shows results with a side length of 5 or less.
The fitting range is adopted to be $0.0 \le t/a^2 \le 2.0$.
Since $\chi (i,j)$ cannot be calculated if the central value of a Wilson loop is negative, $\chi (i,j)$ containing a negative Wilson loop is not used for this fitting.
The string tension is the double limit of $i, j \to \infty$ and $t \to 0$ after removing their finite volume effects, i.e. the string tension is $c(i,j)$ where $(i,j)$ is large.
The results of $c(5,5)$ are $0.248(28)$ from the original Wilson loop and $0.244(14)$ from monopoles.\footnote{
We also fit $\chi (i,j)$ with a quadratic function.
The results of the quadratic fit by the original Wilson loops are
$c(3,3)=0.2814(6)$, $c(4,4)=0.2494(51)$, and $c(5,5)=0.267(66)$.
Those by the monopole Wilson loops are
$c(3,3)=0.2514(3)$, $c(4,4)=0.2358(23)$, and $c(5,5)=0.248(26)$.
The difference between the linear fit and the quadratic fit would be the systematic error due to the choice of fitting function.
}
The Creutz ratio from monopoles is consistent with the original value within the statistical error.

On the other hand, the Creutz ratios computed from $\langle W_f (i,j) \rangle$ are plotted in Fig.~\ref{fig:crephflow}.
Even in the confined phase ($\beta=0.99$), the Wilson loop from field strength does not show the area law, thus the Creutz ratios are zero.
Namely, $F_{\mu \nu}$ becomes smaller by the gradient flow, but originally $F_{\mu \nu}$ does not contribute to the string tension.
The string tension is produced by the monopoles, and the monopoles do not disappear by the gradient flow.

\section{Deconfinement phase transition}
\label{sec:transition}

Next, we discuss the order parameter of the confinement phase transition.
The finite temperature phase transition of $U(1)$ or $SU(N)$ lattice gauge theories without dynamical fermions can be understood as a spontaneous breaking of the center symmetry.
The order parameter for the center symmetry is the Polyakov loop.
The confinement phase is a symmetric phase.
Thus, the expectation value of the Polyakov loop $\langle L \rangle$ is zero in the confinement phase and finite in the deconfinement phase.
We physically interpret the Polyakov loop as $\langle L \rangle \sim e^{-F/T}$, where $F$ is the free energy when there is one charged particle.
Similar to topological stability, there is often symmetry behind the stability of physical quantities.
We study the order parameter: Polyakov loop.

The Polyakov loop operator 
\begin{eqnarray}
L =
\frac{1}{N_s^3} \sum_{\vec{x}} L^{\rm (loc)} (\vec{x})
= \frac{1}{N_s^3} \sum_{\vec{x}} \exp \left[ i \sum_{j=0}^{N_{t}-1} \theta_4 (\vec{x} + j \hat4) \right] ,
\end{eqnarray} 
can be expressed as the product of the contribution $L_m^{\rm (loc)}$ from monopole and the contribution $L_f^{\rm (loc)}$ from field strength \cite{Suzuki:1994ay}. 
We use the identity, 
\begin{eqnarray}
\theta_4 (x)= - \hspace{-1mm} \sum_{x'} \hspace{-1mm} 
D(x-x') \left[ \partial'_{\nu}\Theta_{\nu 4}(x')+
\partial_4 (\partial'_{\nu}\theta_{\nu}(x')) \right] ,
\end{eqnarray}
where $D(x-x')$ is a four-dimensional Coulomb propagator on a lattice satisfying   
$\partial'_{\mu} \partial_{\mu} D(x) = - \delta_{x,0}$, and 
$\Theta_{\mu \nu} = \partial_{\mu} \theta_{\nu} - \partial_{\nu} \theta_{\mu} $.
Then, the local Polyakov loop operator $L^{\rm (loc)} (\vec{x})$ at each point $\vec{x}$ can be decomposed as follows: 
\begin{eqnarray}
L^{\rm (loc)} (\vec{x}) &=& \exp \left[ i \sum_{j=0}^{N_{t}-1} \theta_4 (\vec{x} + j \hat4) \right]
= L_f^{\rm (loc)} (\vec{x}) \cdot L_m^{\rm (loc)} (\vec{x}) , \\ 
L_f^{\rm (loc)} (\vec{x}) &=& \exp \left[ -i\sum_{j=0}^{N_{t}-1} \sum_{x'} D(\vec{x}+j \hat{4} -x') \,
 \partial'_{\nu}\bar{\Theta}_{\nu 4}(x') \right] , 
 \label{eq:plpho} \\
%L_m &=& \exp \left[ -2\pi i\sum_{j=0}^{N_{t}-1} 
% \sum_{x'} D(\vec{x}+ j \hat{4} -x') \, \partial'_{\nu}n_{\nu 4}(x') \right] 
L_m^{\rm (loc)} (\vec{x}) &=& \exp \left[ -2\pi i\sum_{j=0}^{N_{t}-1} \sum_{x'} 
 D(\vec{x}+j \hat{4}-x') \frac{1}{2} \epsilon_{\nu 4 \rho \sigma }
 \partial'_{\nu} \hspace{1mm} ^{\ast} \hspace{-1mm} n_{\rho \sigma}(x') \right] .
\label{eq:plmon}
\end{eqnarray}
Here, 
$\Theta_{\mu\nu}(x) = \bar{\Theta}_{\mu\nu}(x)+ 2\pi n_{\mu\nu}(x), \ 
(-\pi < \bar{\Theta}_{\mu\nu}(x) \le \pi)$. 
$ ^{\ast} \hspace{-1mm} n_{\mu \nu} (x) = 
\frac{1}{2} \epsilon_{\mu \nu \rho \sigma} n_{\rho \sigma}(x) $ 
is the Dirac string explained in Sec.~\ref{sec:monopole}.
The magnetic monopole currents $k_{\mu}(x)$ are the boundary of the Dirac string (sheet), i.e. 
$k_{\mu} (x) = \partial_{\nu} \: ^{\ast} \hspace{-1mm} n_{\nu \mu} (x)$.
We denote the spatial averages of $L_f^{\rm (loc)} (\vec{x})$ and $L_m^{\rm (loc)} (\vec{x})$ in Eqs.~(\ref{eq:plpho}) and (\ref{eq:plmon}) as $L_f$ and $L_m$.
The expectation values $\langle L_f \rangle$ and $\langle L_m \rangle$ are the contributions from $F_{\mu \nu}$ and monopoles to $\langle L \rangle$, respectively. 
The contribution from monopoles $\langle L_m \rangle$ is calculated by the Dirac string, not from the monopole current.
However, as explained below, except for the plus and minus signs, the value of the Polyakov loop is determined solely by the location of the monopole current, not the Dirac string (sheet).
Therefore, we call $L_m$ the contribution from monopoles.

In order to consider the behavior of the Polyakov loop in relation to the configuration change of the monopole currents, we rewrite $L_m^{\rm (loc)} (\vec{x})$.
%Using the Dirac string  
%$^{\ast} \hspace{-1mm} n_{\mu \nu} (x) = 
%\frac{1}{2} \epsilon_{\mu \nu \rho \sigma} n_{\rho \sigma}(x)$, 
%$L_m$ becomes 
%\begin{eqnarray}
%L_m = \exp \left[ -2\pi i\sum_{i=0}^{N_{t}-1} \sum_{x'} 
% D(\vec{x}+i \hat{4}-x') \frac{1}{2} \epsilon_{\nu 4 \rho \sigma }
% \partial'_{\nu} \hspace{1mm} ^{\ast} \hspace{-1mm} n_{\rho \sigma}(x') \right] .
%\label{eq:pmon}
%\end{eqnarray}
Integrating out the time direction, 
\begin{eqnarray}
%L_m &=& \exp \left[ -2\pi i \sum_{\vec{x}'} \partial'_{i}
% \sum_{x_{4}=1}^{N_{t}} 
% D(\vec{x}-\vec{x}', x_{4}) \frac{1}{2} \epsilon_{i j k 4}
% \sum_{x'_{4}=1}^{N_{t}} \hspace{1mm} ^{\ast} \hspace{-1mm} 
% n_{j k}(\vec{x}',x'_{4}) \right]  \nonumber \\
L_m^{\rm (loc)} (\vec{x}) &=& \exp \left[ -2\pi i \sum_{\vec{x}'} \partial'_{i} D_{3}(\vec{x}-\vec{x}') 
 \frac{1}{2} \epsilon_{i j k 4} \hspace{1mm} ^{\ast} \hspace{-1mm} 
 \tilde{n}_{j k}(\vec{x}',x'_{4}) \right] , 
\label{eq:p3mon}
\end{eqnarray}
where $\vec{x}$ is the position of the Polyakov loop in three-dimensional space, and
\begin{eqnarray}
D_{3}(\vec{x}) = \sum_{x_{4} =1}^{N_{t}} D(\vec{x},x_{4}) , 
\end{eqnarray} 
is a three-dimensional Coulomb propagator, since it satisfies 
\begin{eqnarray}
\partial'_{i} \partial_{i} D_{3}(\vec{x}) = 
\sum_{x_{4}} \left[ \partial'_{i} \partial_{i} D(\vec{x},x_{4}) + 
\partial'_{4} \partial_{4} D(\vec{x},x_{4}) \right] = - \delta_{\vec{x},0}.
\end{eqnarray}
$^{\ast} \hspace{-1mm} \tilde{n}_{i j} (\vec{x})$ is a projection of Dirac string in three-dimensional space, 
\begin{eqnarray}
^{\ast} \hspace{-1mm} \tilde{n}_{i j} (\vec{x}) =
\sum_{x_{4}=1}^{N_{t}}\hspace{1mm} ^{\ast} \hspace{-1mm} n_{i j} (\vec{x},x_{4}).
\end{eqnarray}
In addition, we introduce the solid angle $\Omega (\vec{x})$ when we see the Dirac string (sheet) from the point $\vec{x}$ where the Polyakov loop is placed \cite{Ejiri:1996sz,Ejiri:1998xf}.
Then, Eq.~(\ref{eq:p3mon}) is rewritten as
\begin{eqnarray}
L_m^{\rm (loc)} (\vec{x}) = \exp \left[ \frac{2\pi i \Omega (\vec{x})}{4 \pi} \right] . 
\label{eq:psang}
\end{eqnarray}
The value of 
$\frac{1}{2} \epsilon_{i j k 4} \hspace{1mm} ^{\ast} \hspace{-1mm} \tilde{n}_{j k}(\vec{x}',x'_{4})$ is the area of the Dirac sheet, whose direction is perpendicular to the plane.
In the continuum theory, $\partial'_{i} D_{3}(\vec{x}-\vec{x}')$ has a magnitude of $1/r^2$, where $r$ is the distance between $\vec{x}$ and $\vec{x}'$, 
and its direction is the direction of looking at the Dirac sheet from $\vec{x}$.
Therefore, the inner product of these quantities is a solid angle $\Omega (\vec{x})$ in three-dimensional space.

Note that this solid angle is signed according to the direction of the monopole current.
From this equation, it is found that the Polyakov loop depends only on the location of the monopole current (boundary of the sheet), not on the shape of the Dirac sheet.
Even if the shape of the Dirac sheet is changed, the value of $L_m^{\rm (loc)} (\vec{x})$ does not change.
Especially, if there is a Dirac sheet bubble without a monopole, i.e. a closed curved Dirac sheet with no boundaries, the solid angle $\Omega (\vec{x})$ looking at the Dirac sheet will be $\pm 4 \pi$ inside the bubble and $0$ outside the bubble.
Therefore, such bubbles do not affect $L_m$. 
However, the sign of the Polyakov loop may change depending on the Dirac sheet.
If there is an infinitely wide Dirac sheet without monopoles, half of the full solid angle, $\pm 2 \pi$, will be added to $\Omega (\vec{x})$, regardless of which $\vec{x}$ we look at the sheet from.
Then, the sign of $L_m^{\rm (loc)} (\vec{x})$ changes at all $\vec{x}$.
Thus, the monopole part of the Polyakov loop operator $L_m$ is a quantity determined only by the distribution of monopoles except for its sign.

From Eq.~(\ref{eq:psang}), when the monopoles are distributed throughout the space, the solid angle $\Omega (\vec{x})$ takes a completely random value for each Polyakov loop location $\vec{x}$.
Then, the spatial average $L_m$ becomes zero; 
whereas, when there are few monopoles, the solid angle is almost zero and the spatial average is close to one.
Furthermore, since the solid angle has a plus or minus sign depending on the direction, even if there are many short closed monopole current loops, they do not contribute much to $\Omega (\vec{x})$.
Equation~(\ref{eq:psang}) suggests that long monopole loops are important because the monopole current spread over space gives $\Omega (\vec{x})$ various values from $-2 \pi$ to $2 \pi$.\footnote{
The importance of long monopole currents in the finite temperature phase transition of Abelian projected QCD is pointed out in Ref.~\cite{Ejiri:1994uw}.
}

\subsection*{Numerical simulation}

\begin{figure}[tb]
\begin{center}
\vspace{0mm}
\includegraphics[width=8.1cm]{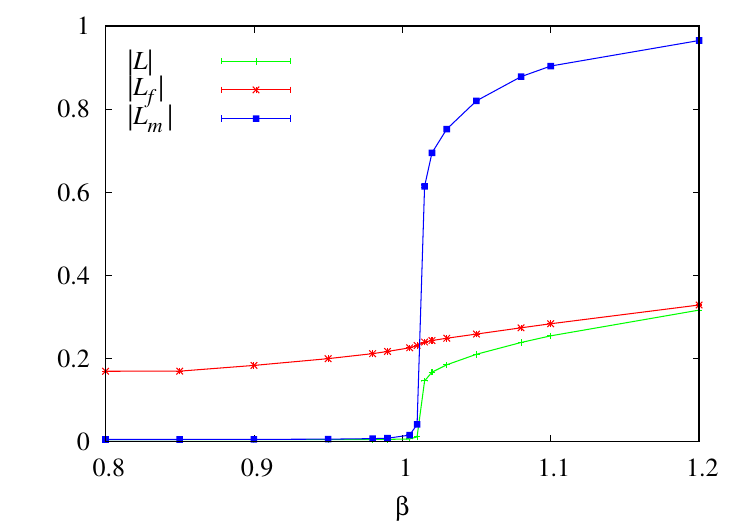}
\hspace{0mm}
\includegraphics[width=8.1cm]{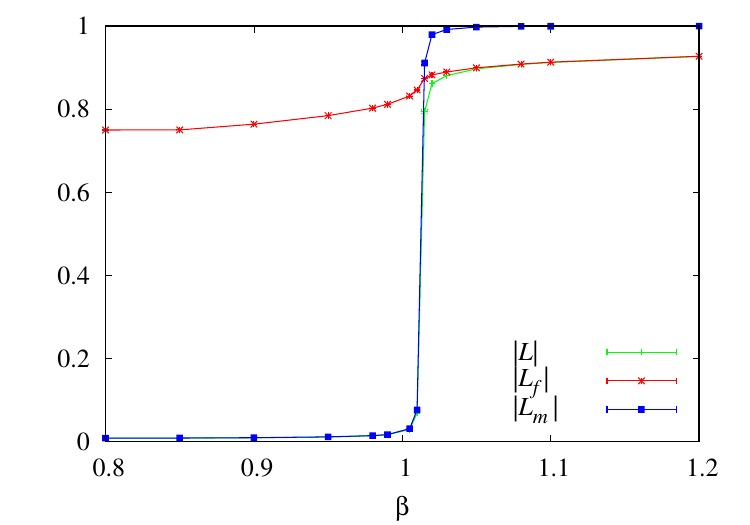}
\vspace{-11mm}
\end{center}
\caption{Absolute value of Polyakov loop as a function of $\beta$ measured on a $32^3 \times 8$ lattice. 
Green, red and blue lines are the original, field strength part and monopole part, respectively.
The left panel is before the gradient flow, and the right panel is the result at $t/a^2 =2.0.$
}
\label{fig:pl-b}
\end{figure}

\begin{figure}[tb]
\begin{center}
\vspace{0mm}
\includegraphics[width=8.1cm]{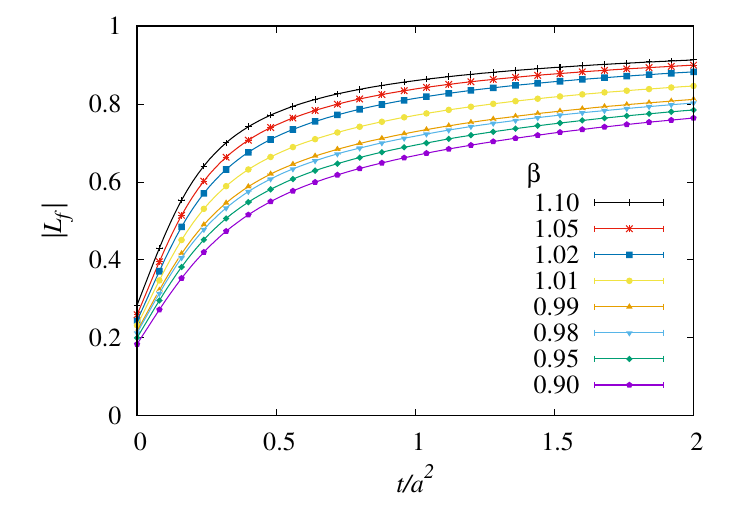}
\hspace{0mm}
\includegraphics[width=8.1cm]{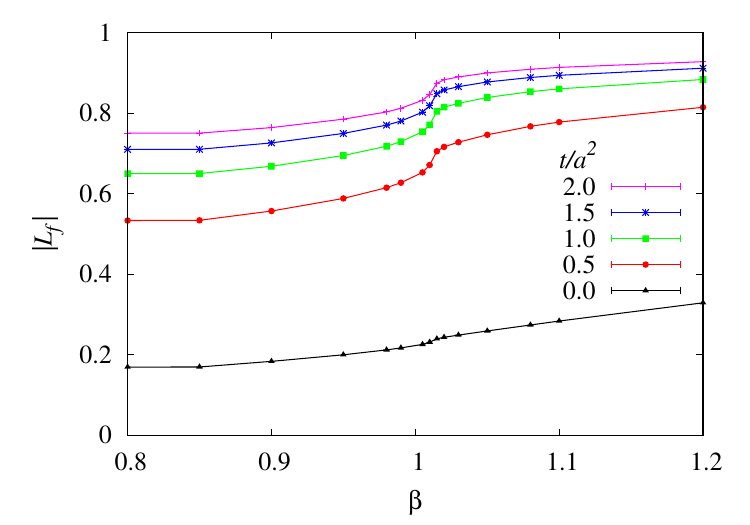}
\vspace{-11mm}
\end{center}
\caption{The left panel is the absolute values of Polyakov loops from field strength as a function of flow time $t/a^2$ for each $\beta$ computed on a $32^3 \times 8$ lattice. 
The right panel is that as a function of $\beta$ at various $t/a^2$.
}
\label{fig:plmonflow}
\end{figure}

\begin{figure}[tb]
\begin{center}
\vspace{0mm}
\includegraphics[width=8.1cm]{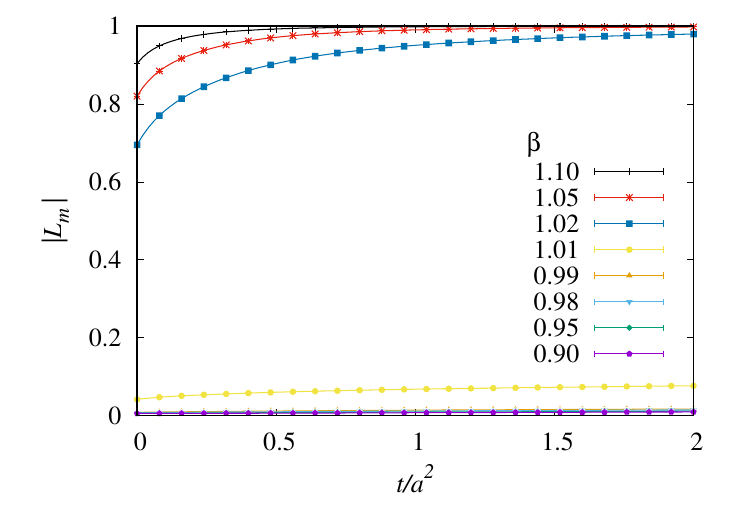}
\hspace{0mm}
\includegraphics[width=8.1cm]{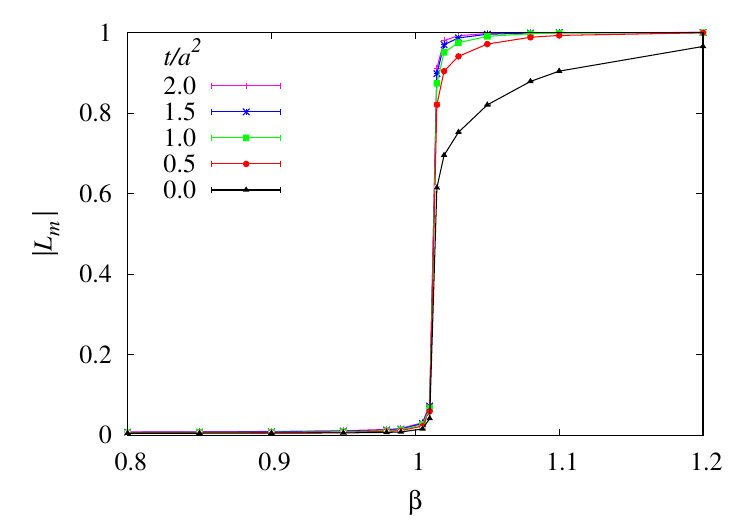}
\vspace{-11mm}
\end{center}
\caption{The left panel is the absolute values of Polyakov loops from monopoles as a function of flow time $t/a^2$ for each $\beta$ computed on a $32^3 \times 8$ lattice. 
The right panel is that as a function of $\beta$ at various $t/a^2$.
}
\label{fig:plmonb}
\end{figure}

\begin{figure}[tb]
\begin{center}
\vspace{0mm}
\includegraphics[width=8.1cm]{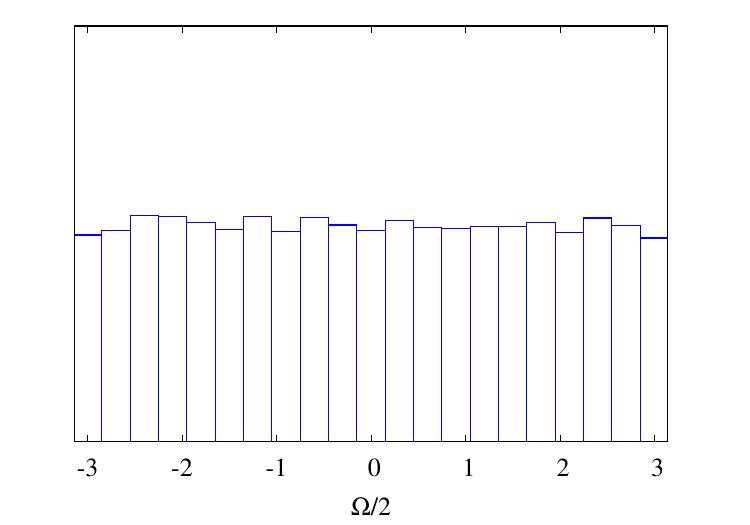}
\hspace{0mm}
\includegraphics[width=8.1cm]{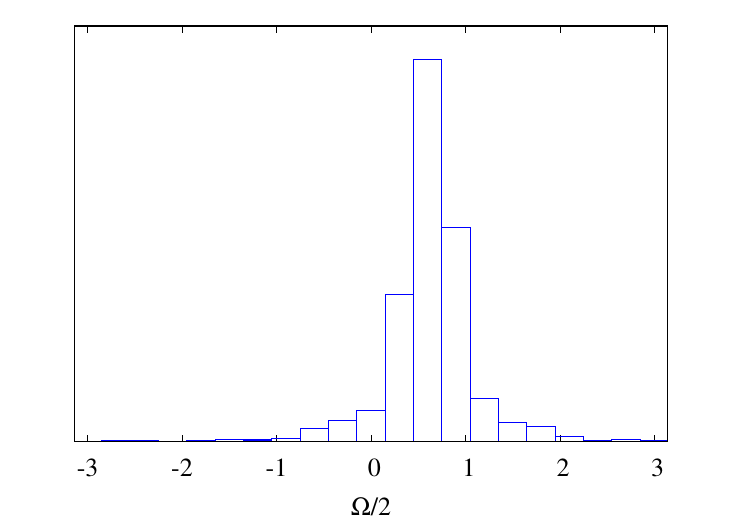}
\vspace{-11mm}
\end{center}
\caption{Histogram of half the solid angle in Eq.~(\ref{eq:psang}).
The left and right panels are the results for the confinement phase $(\beta=0.98)$ and the deconfinement phase $(\beta=1.10)$ on a $32^3 \times 8$ lattice, respectively.
}
\label{fig:sangpbc}
\end{figure}

We calculate the Polyakov loop separating the monopole contribution $L_m$ and the field strength contribution $L_f$ by performing Monte Carlo simulations.
The lattice size is $32^3 \times 8$ and the number of configurations is 2000, which is the same as Sec.~\ref{sec:gradient}.
The $\beta$ dependence of the absolute values of Polyakov loops, $\langle | L | \rangle$, $\langle | L_f | \rangle$, and $\langle | L_m | \rangle$ is shown in Fig.~\ref{fig:pl-b}.
The green, red and blue lines are the expectation value of the original one, field strength part and monopole part, respectively.
The left panel presents the result before the gradient flow and the right one the result at flow time $t/a^2=2.0$.
The monopole contribution in the confinement phase below $\beta \approx 1.01$ is zero, and remains zero even when flowed.
Then, the monopole part increases sharply at the phase transition point, and the increase is steeper after the gradient flow.
On the other hand, the field strength contribution is nonzero at all $\beta$ and increases with the gradient flow.
It is found that the contribution from monopoles makes the Polyakov loop zero in the confinement phase and the field strength part does not contribute to the deconfinement phase transition.

We also plot the field strength contribution and the monopole contribution as functions of the flow time for each $\beta$ and functions of $\beta$ for each $t/a^2$ in Figs.~\ref{fig:plmonflow} and \ref{fig:plmonb}.
Figure~\ref{fig:plmonflow} is the result of field strength part $\langle | L_f | \rangle$. 
As shown in the left panel, $\langle | L_f | \rangle$ monotonically increases with increasing the flow time, and the dependence on $\beta$ is small as in the right panel.
Figure~\ref{fig:plmonb} is the monopole part $\langle | L_m | \rangle$.
$\langle | L_m | \rangle$ remains zero in the confinement phase and approaches one in the deconfinement phase as the flow time increases in the left panel.
Then, as seen in the right panel, $\langle | L_m | \rangle$ after the gradient flow becomes like a step function in $\beta$.

In addition, histograms of the solid angle $\Omega (\vec{x})$ in Eq.~(\ref{eq:psang}) are shown in Fig.~\ref{fig:sangpbc}.
These are the histogram of $\Omega (\vec{x})$ at all spatial points $\vec{x}$ on one configuration measured on  the $32^3 \times 8$ lattice before the gradient flow. 
The left panel is the results at $\beta=0.98$ in the confinement phase.
The distribution probability in terms of $\Omega (\vec{x})$ is almost constant.
Then, the spatial average of $L_m^{\rm (loc)} (\vec{x})$ is approximately zero.
On the other hand, the right panels are the results at $\beta=1.10$ in the deconfinement phase.
The histogram in the deconfinement phase is shaped with a sharp peak.
Then, the spatial average of $L_m^{\rm (loc)} (\vec{x})$ is nonzero.
However, the discussion in this section suggests that the histogram of $\Omega (\vec{x})$ has a peak at $\Omega =0$ when there are few monopoles in the deconfinement phase.
As seen in the right panel of Fig.~\ref{fig:sangpbc}, the peak position is nonzero in practical calculations.
This peak shift will be discussed in the next section.

\section{Center symmetry breaking}
\label{sec:center}

\subsection{Order parameter of the center symmetry}

We consider the following transformation of $U_{4}(x)$ in the $U(1)$ lattice gauge theory, called the $U(1)$ center transformation, 
\begin{eqnarray}
U_{4}(\vec{x},t_1) \longrightarrow e^{i \phi} U_{4}(\vec{x},t_1) ,
\end{eqnarray}
at all spatial points $\vec{x}$ in one time slice $t_1$.
$\phi$ is an arbitrary real number.
Then, $\theta_{4}(\vec{x},t_1) \to \theta_{4}(\vec{x},t_1) + \phi$ in Mod $2\pi$.
Because $e^{i \phi}$ is an element of the $U(1)$ group, the path integral measure ${\cal D} U_\mu (x)$ is invariant. 
Since 
$\cos \Theta_{\mu \nu}(x)= \mathrm{Re}[ U_{\mu}(x) U_{\nu}(x+\hat{\mu}) 
U^*_{\mu}(x+\hat{\nu}) U^*_{\nu}(x)]$
does not change under the center transformation, the action does not change.
However, the Polyakov loop changes from $\langle L \rangle$ to $e^{i \phi} \langle L \rangle$ because the loop passes through the time slice $t_1$ only once.
Therefore, the expectation value of the Polyakov loop is zero unless the center symmetry is spontaneously broken.
In other words, the phase transition in which the Polyakov loop changes from zero to a finite value is a phase transition due to the spontaneous breaking of this global $U(1)$ center symmetry.
The Polyakov loop is an order parameter for the $U(1)$ center symmetry breaking.

We interpret the Polyakov loop as $\langle L \rangle \sim e^{-F/T}$, where $F$ is the free energy when there is one charged particle.
For this interpretation, the $U(1)$ symmetric complex phase of the Polyakov loop is unphysical that must be removed.
The $U(1)$ symmetry comes from the extra degree of freedom in the theory.
In the context of monopole condensations, it is interesting to consider which extra dynamic variables are related to the $U(1)$ center symmetry.
In the following, we focus on the Dirac string, which is an unphysical unobserved quantity in Dirac's magnetic monopole theory.

\begin{figure}[tb]
\begin{center}
\vspace{0mm}
\includegraphics[width=8.1cm]{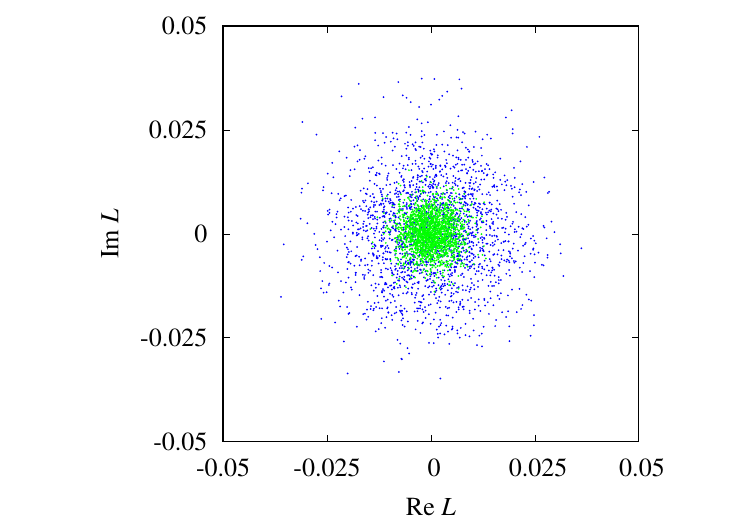}
\hspace{0mm}
\includegraphics[width=8.1cm]{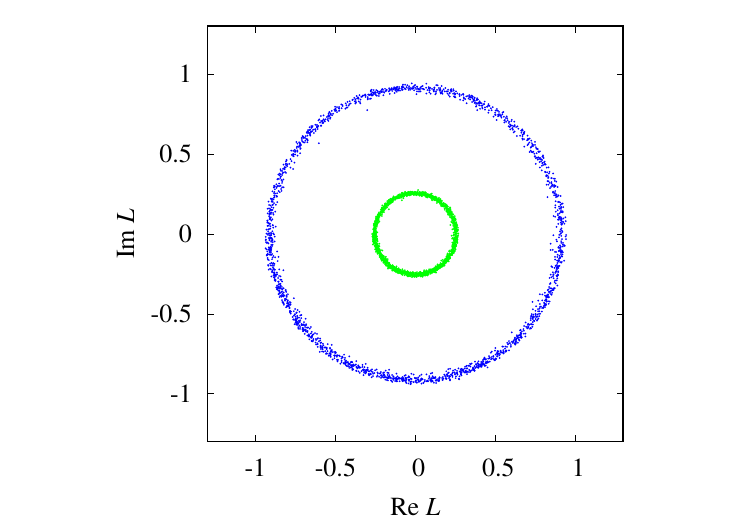}
\vspace{-11mm}
\end{center}
\caption{Distribution of Polyakov loops in the complex plane.  
The left and right panels are the results $\beta=0.98$ and 1.10 on a $32^3 \times 8$ lattice, respectively.
Green symbols are before the gradient flow, and blue symbols are at $t/a^2 =2.0$.
}
\label{fig:pldis}
\end{figure}

\begin{figure}[tb]
\begin{center}
\vspace{0mm}
\includegraphics[width=8.1cm]{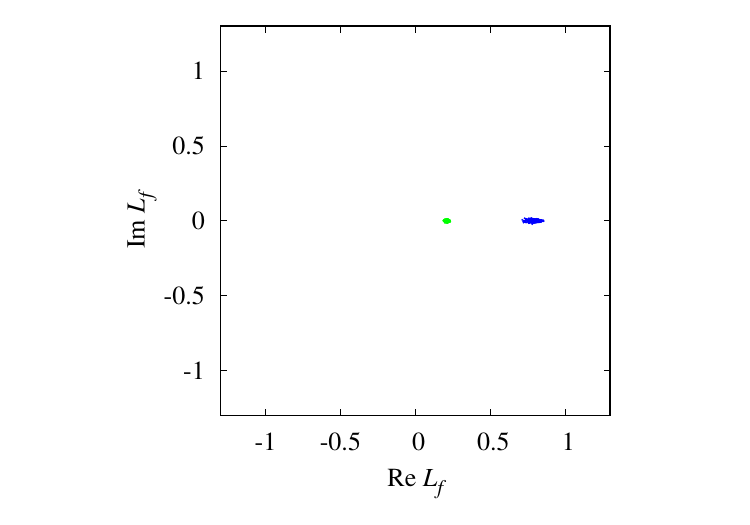}
\hspace{0mm}
\includegraphics[width=8.1cm]{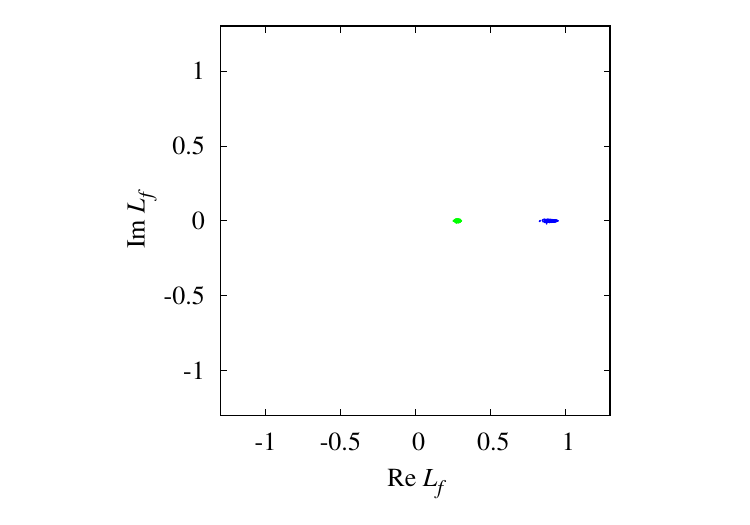}
\vspace{-11mm}
\end{center}
\caption{The distribution of the field strength part of the Polyakov loop in the complex plane at $\beta=0.98$ (left) and 1.10 (right) before the gradient flow (green) and at $t/a^2=2.0$ (blue).
}
\label{fig:plphdis}
\end{figure}

\begin{figure}[tb]
\begin{center}
\vspace{0mm}
\includegraphics[width=8.1cm]{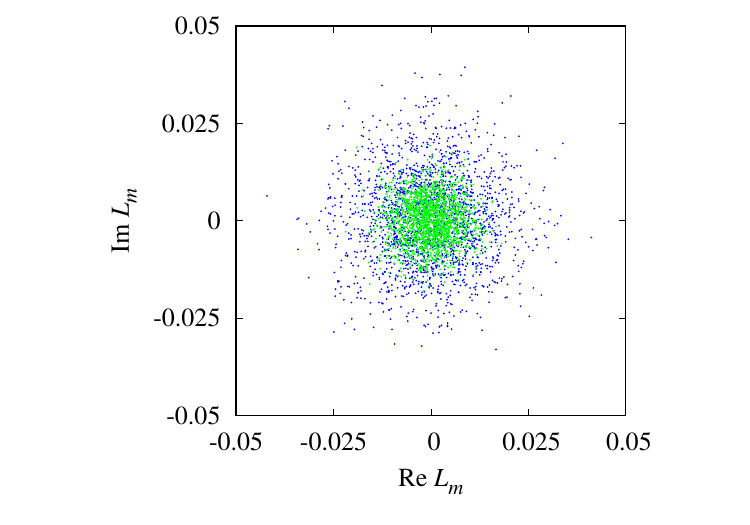}
\hspace{0mm}
\includegraphics[width=8.1cm]{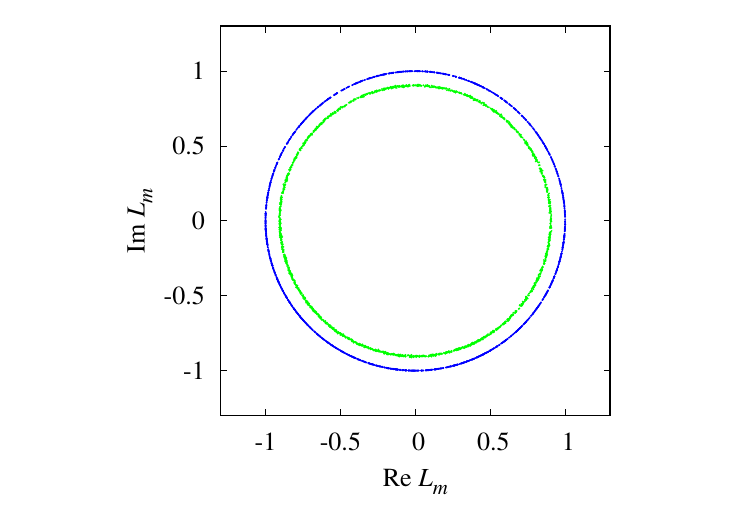}
\vspace{-11mm}
\end{center}
\caption{The distribution of the monopole part of the Polyakov loop in the complex plane at $\beta=0.98$ (left) and 1.10 (right) before the gradient flow (green) and at $t/a^2=2.0$ (blue).}
\label{fig:plmondis}
\end{figure}

Since the action and the path integral measure are invariant under the $U (1)$ center transformation, the distribution probabilities at $L$ and $e^{i \phi} L$ are equal for any $\phi$.
When the Polyakov loop values on each configuration are plotted on the complex plane, the distribution must be $U(1)$ symmetric.
The left and right panels of Fig.~\ref{fig:pldis} are the results the Polyakov loops averaged over the space at $\beta=0.98$ (confinement) and $1.10$ (deconfinement), respectively.
The green dots are the result before the gradient flow and the blue dots are that at $t/a^2=2.0$.
The distribution is $U(1)$ symmetric in both the confinement and deconfinement phases.
These are distributed near the origin in the confinement phase and on a circle in the deconfinement phase.
Spontaneous symmetry breaking occurs when one of the points on the circle with the highest probability is chosen as the vacuum in the deconfinement phase.
Therefore, when the Polyakov loop is decomposed into $L_f$ and $L_m$, the distribution of the part related to the spontaneous symmetry breaking must be $U(1)$ symmetric.

We expect that the field strength part $L_f$ is unrelated to the center symmetry, since $\bar{\Theta}_{\mu \nu}(x)$ does not change under the center transformation.
The distribution of the field strength part $L_f$ on the complex plane is shown in Fig.~\ref{fig:plphdis}.
The green and blue dots are the results before and after the gradient flow, respectively.
These figures show that the distribution is not $U(1)$ symmetric and the field strength part is not related to the center symmetry. 

We also plot the distribution of the monopole part $L_m$ on the complex plane in Fig.~\ref{fig:plmondis}.
From these figures, the distributions of the monopole part are found to be $U(1)$ symmetric both before and after the gradient flow, similar to the original $L$.
This result indicates that the important part for the center symmetry breaking is the monopole part.
However, it is seemingly incomprehensible that the monopole part is $U(1)$ symmetric because the monopole current is invariant under the central transformation.
[See the definition of the magnetic monopole current Eq.~(\ref{eq:monopole}).]
Furthermore, in the deconfinement phase, the number of monopoles is reduced and the complex phase of $L_m$ should take a value near zero.
The origin of the symmetric complex phase of $L_m$ will be discussed in the following subsections.

Moreover, this situation does not change even if the gauge field is coarse grained by the gradient flow equation considering the compactness Eq.~(\ref{eq:coflow}) as seen in Figs.~\ref{fig:pldis} and \ref{fig:plmondis}.
The reason that the gradient flow does not break this $U(1)$ symmetry is that the flow equation is invariant under the center transformation.
On the other hand, the flow equation which does not consider compactness, Eq.~(\ref{eq:noncoflow}), does not have the $U(1)$ center symmetry. 
When Eq.~(\ref{eq:noncoflow}) is used, coarse graining with the gradient flow immediately breaks the $U(1)$ symmetry of the Polyakov loop distribution.
Then, the nature of confinement is lost by the gradient flow as we have seen in Sec.~\ref{sec:noncompact}.

\subsection{Coulomb propagator and boundary condition}

Before discussing the center symmetry with respect to $L_m$, we need to discuss the Coulomb propagator on a finite lattice.
So far, the discussion has been on a system with infinite volume, but the case of finite volume with periodic boundary conditions is not so simple.
Strictly speaking, when the periodic boundary condition 
\begin{eqnarray}
D(x) = D(x + N_s \hat{\mu}) 
\end{eqnarray}
is imposed for $\mu$ direction with $\mu=1, 2, 3, 4$, the definition of the Coulomb propagator, 
$\partial'_{\mu} \partial_{\mu} D(x-y) = - \delta_{x,y}$, 
cannot be satisfied.
When both sides are integrated for the entire space, the right-hand side becomes $-1$.
However, since the left-hand side is the total derivative, and the boundary value is canceled by the periodic boundary condition, the left-hand side becomes zero.
This is because the operator $\partial'_{\mu} \partial_{\mu}$ has zero eigenvalues.
The eigenvector $\vec{v}$ of the zero eigenvalue is when all of the components are the same, i.e. $v_x=c$, where $c$ is an arbitrary constant because $\partial'_{\mu} \partial_{\mu} v_x =0$.
Since $D (x-y)$ is the inverse matrix of $\partial'_{\mu} \partial_{\mu}$, $D (x-y)$ does not exist. 

In such cases, the definition should be changed to satisfy the propagator definition only when multiplied by any vector except the eigenvectors of zero eigenvalues $\vec{v}$. 
A projection operator on the vector spaces excluding zero eigenvalues can be defined as follows, 
\begin{eqnarray}
\mathcal{P}_{xy} = \delta_{x,y} - \frac{v_x v_y}{|\vec{v}|^2},
\end{eqnarray}
which satisfies $\mathcal{P} \vec{v} =0$.
Using the projection operator, the propagator satisfies the equation,
\begin{eqnarray}
\partial'_{\mu} \partial_{\mu} D(x-y) \mathcal{P}_{yz} \psi_z = -\mathcal{P}_{xy} \psi_y,
\label{eq:Dproj}
\end{eqnarray}
where $\vec{\psi}$ is an arbitrary vector, and $\mathcal{P} \vec{\psi}$ is a vector excluding zero eigenvalues.
The following equation satisfies Eq.~(\ref{eq:Dproj}) for any $\vec{\psi}$ because $\mathcal{P} \mathcal{P} =\mathcal{P}$, 
\begin{eqnarray}
\partial'_{\mu} \partial_{\mu} D(x-y) = -\mathcal{P}_{xy} .
%= -\delta_{x,y} + \frac{v_x v_y}{|\vec{v}|^2}.
\end{eqnarray}
For the case of $v_x=c$,  
\begin{eqnarray}
\mathcal{P}_{xy} = \delta_{x,y} - \frac{cc}{N_{\rm site} c^2} 
= \delta_{x,y} - \frac{1}{N_{\rm site}},
\end{eqnarray}
and the definition of the Coulomb propagator becomes
\begin{eqnarray}
\partial'_{\mu} \partial_{\mu} D(x-y) = \delta_{x,y} - \frac{1}{N_{\rm site}}.
\label{eq:cpropp}
\end{eqnarray}
This definition is the same as removing the zero-momentum mode that diverges when one calculates $D(x)$ by the Fourier transform.

\subsection{Additional complex phase in the Polyakov loop}

In this definition of the Coulomb propagator, for example, a bubble of Dirac sheet defined in a unit cube at a point $y$ shifts the phase of the Polyakov loop by $2\pi/N_s^3$.
We substitute 
\begin{eqnarray}
^{\ast} \hspace{-1mm} n_{12}(y)
= \, ^{\ast} \hspace{-1mm} n_{23}(y)
= \, ^{\ast} \hspace{-1mm} n_{31}(y)=1, 
\hspace{3mm} 
^{\ast} \hspace{-1mm} n_{12}(y-\hat{3})
= \, ^{\ast} \hspace{-1mm} n_{23}(y-\hat{1})
= \, ^{\ast} \hspace{-1mm} n_{31}(y-\hat{2}) = -1
\label{eq:unitcube}
\end{eqnarray}
into Eq.~(\ref{eq:plmon}). 
The Dirac sheet is antisymmetric with respect to the directional index and 
$^{\ast} \hspace{-1mm} n_{\rho \sigma}(x')=0$ elsewhere. 
Note that there is no monopole current in this case.
Also see footnote~\ref{footnote3} to understand that Eq.~(\ref{eq:unitcube}) means a unit cubic bubble.
Then, 
\begin{eqnarray}
L_m^{\rm (loc)} (\vec{x}) &=& \exp \left[-2\pi i\sum_{j=0}^{N_{t}-1} \sum_{k=1}^{3} [ 2D(\vec{x}+j \hat{4}-y) 
- D(\vec{x}+j \hat{4}-y+\hat{k}) - D(\vec{x}+j \hat{4}-y-\hat{k}) ] \right] \nonumber \\
% &=& \exp \left[-2\pi i\sum_{j=0}^{N_{t}-1} \sum_{k=1}^{3} [ \partial'_{k} D(\vec{x}+j \hat{4}-y) 
%- \partial'_{k} D(\vec{x}+j \hat{4}-y+\hat{k} ] \right] \nonumber \\
&=& \exp \left[ 2\pi i\sum_{j=0}^{N_{t}-1} \partial'_{k} \partial_{k} D(\vec{x}+j \hat{4}-y) \right] 
% = \exp \left[ 2\pi i\sum_{j=0}^{N_{t}-1} \partial'_{\mu} \partial_{\mu} D(\vec{x}+j \hat{4}-y) \right] 
= \exp \left[ 2\pi i\sum_{j=0}^{N_{t}-1} \left( -\delta_{\vec{x}+j \hat{4}, y} + \frac{1}{N_{\rm site}} \right) \right] 
\nonumber \\
&=& \exp \left( \frac{2\pi i}{N_s^3} \right) .
\label{eq:bubble}
\end{eqnarray}
Thus, extra complex phases appear without monopoles.
Regardless of the location of the bubble, the phase of $2\pi/N_s^3$ is added to the phase of the Polyakov loop $L_m^{\rm (loc)} (\vec{x})$ and also $L_m$ for the unit bubble of Dirac sheet.
Therefore, when there is a bubble of Dirac sheet of size $N_{\rm bubble}$, the Polyakov loop is multiplied by the phase $\exp(2\pi i N_{\rm bubble}/N_s^3)$ \cite{Ejiri:1996sz,Ejiri:1998xf}.
The origin of this additional phase is the exclusion of the constant mode in the Coulomb propagator.
We discuss in Appendix \ref{sec:antiprio} the case of imposing an antiperiodic boundary conditions such that the definition of Coulomb propagator Eq.~(\ref{eq:cprop}) holds strictly.
In that case, no extra phase is added to the Polyakov loop other than the plus or minus sign.
However at the same time, the center symmetry is reduced from $U(1)$ to $Z_2$.

In the $U(1)$ lattice gauge theory, the theory is invariant under the transformation of adding a Dirac sheet without monopoles, but the phase of the Polyakov loop changes according to the size of the Dirac sheet bubbles.
This corresponds to the change in complex phase associated with the $U(1)$ center transformation.
Under the center transformation, the monopole current never changes, but the Dirac string changes.
Then, the size of the Dirac sheet bubbles of $N_{\rm bubble}$ changes, and the additional complex phase $\exp(2\pi i N_{\rm bubble}/N_s^3)$ changes.
This phase shift is discrete, but can be regarded as a continuous change if the spatial volume is large enough.
Thus, the center symmetry is reflected in the symmetry of the distribution of the monopole part of the Polyakov loop $L_m$.
In addition, it is important to perform the gradient flow so as not to break the symmetry to keep the nature of the confinement.
The flow equation Eq.~(\ref{eq:coflow}) is symmetric under the center transformation.

\subsection{Gradient flow and Polyakov loops}

\begin{figure}[tb]
\begin{center}
\vspace{0mm}
\includegraphics[width=8.1cm]{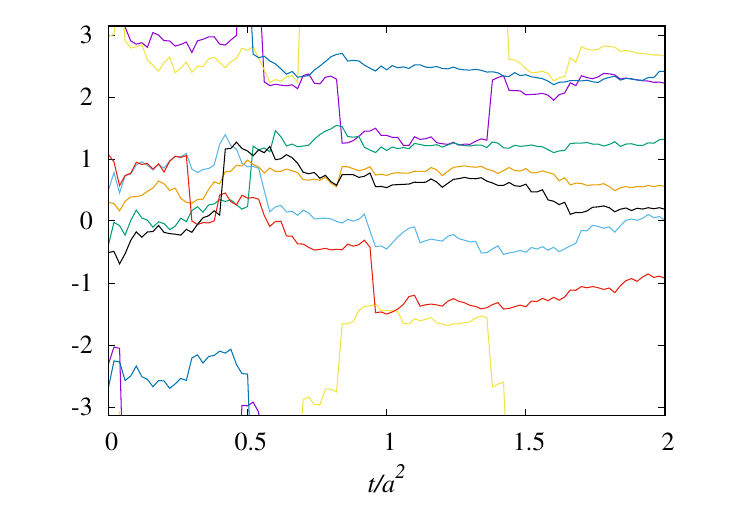}
\hspace{0mm}
\includegraphics[width=8.1cm]{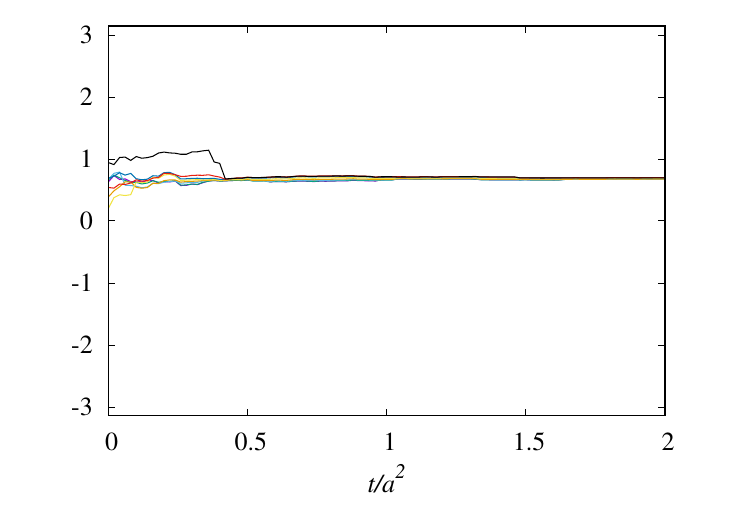}
\vspace{-11mm}
\end{center}
\caption{Phases of the monopole part of Polyakov loops $\Omega (\vec{x})/2$ at some points on one configuration. The left and right panels are the results at $\beta=0.98$ and $1.10$, respectively.}
\label{fig:lplphasemo}
\end{figure}

\begin{figure}[tb]
\begin{center}
\vspace{0mm}
\includegraphics[width=8.1cm]{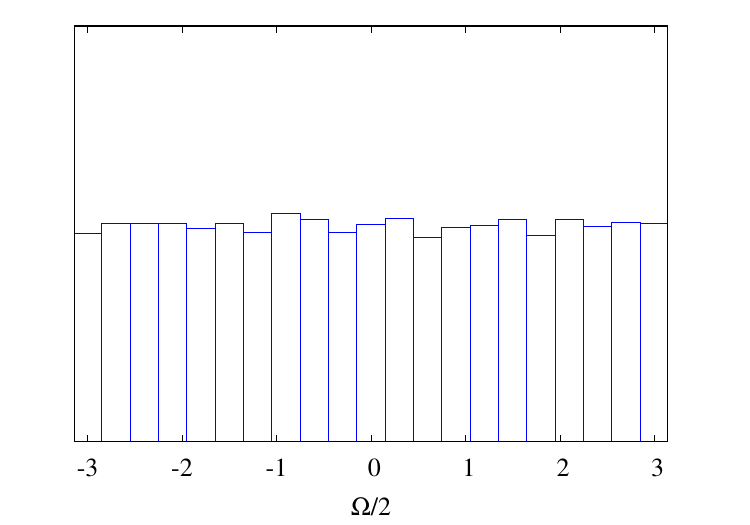}
\hspace{0mm}
\includegraphics[width=8.1cm]{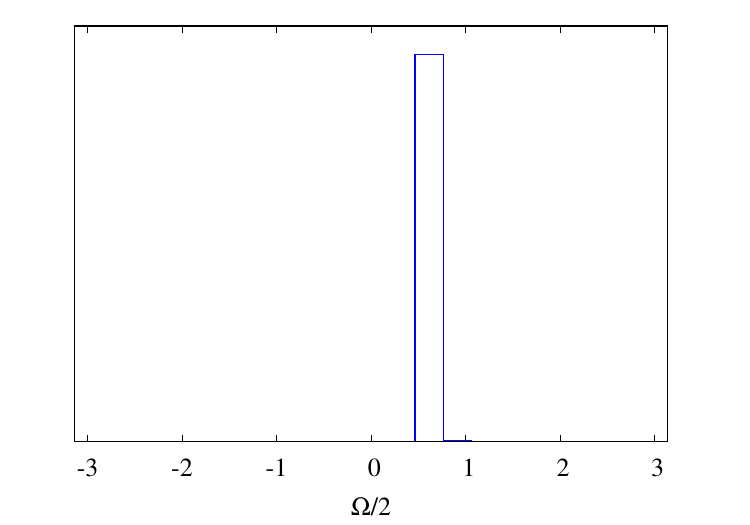}
\vspace{-11mm}
\end{center}
\caption{Histogram of half the solid angle in Eq.~(\ref{eq:psang}) after the gradient flow.
The left and right panels are the results for the confinement phase $(\beta=0.98)$ and the deconfinement phase $(\beta=1.10)$ at $t/a^2=2.0$ on a $32^3 \times 8$ lattice, respectively.
}
\label{fig:sangpbct2}
\end{figure}

\begin{figure}[tb]
\begin{center}
\vspace{0mm}
\includegraphics[width=8.1cm]{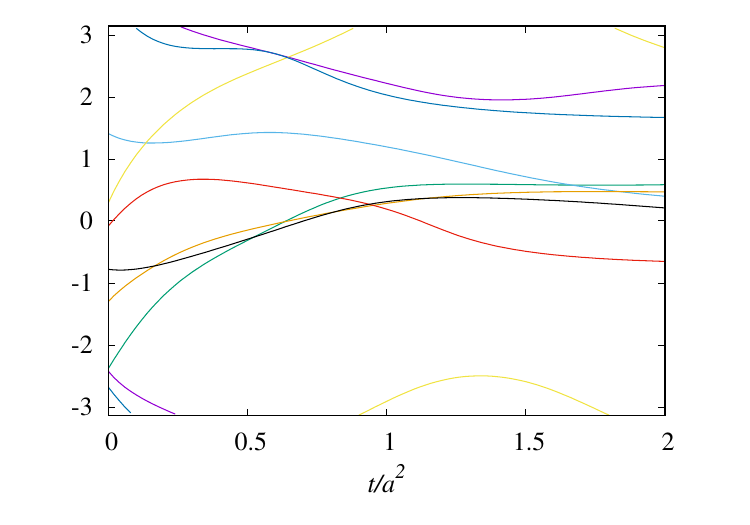}
\hspace{0mm}
\includegraphics[width=8.1cm]{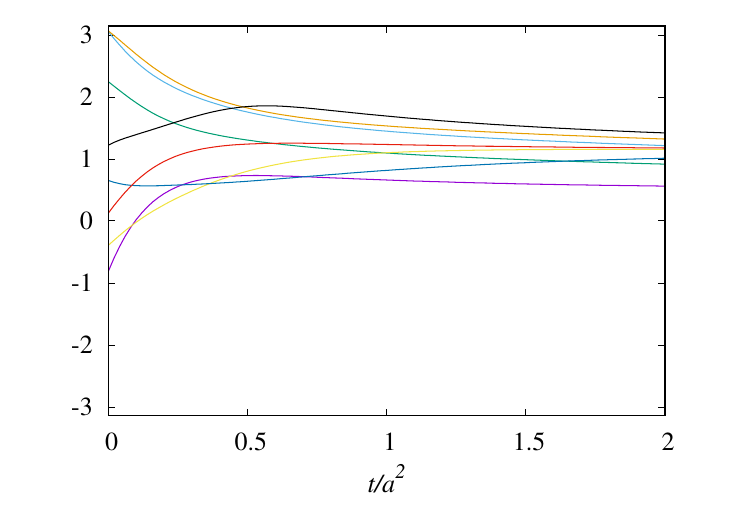}
\vspace{-11mm}
\end{center}
\caption{Phases of Polyakov loops at some points on one configuration. 
The left and right panels are the results at $\beta=0.98$ and $1.10$, respectively.}
\label{fig:lplphase}
\end{figure}

\begin{figure}[tb]
\begin{center}
\vspace{0mm}
\includegraphics[width=8.1cm]{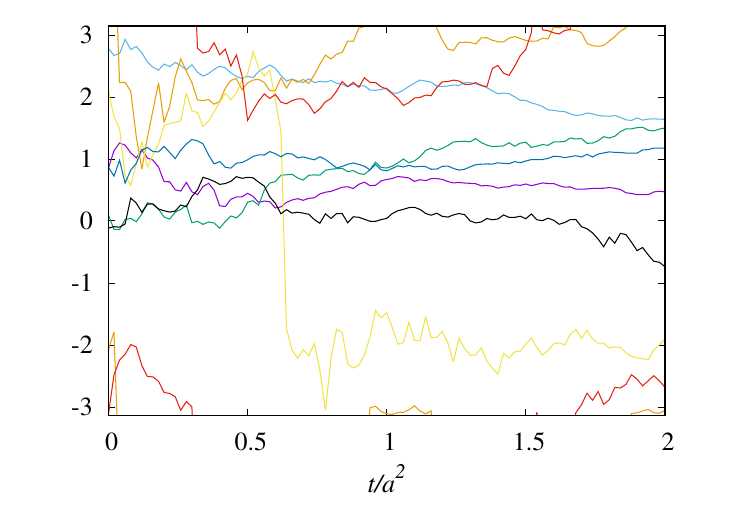}
\hspace{0mm}
\includegraphics[width=8.1cm]{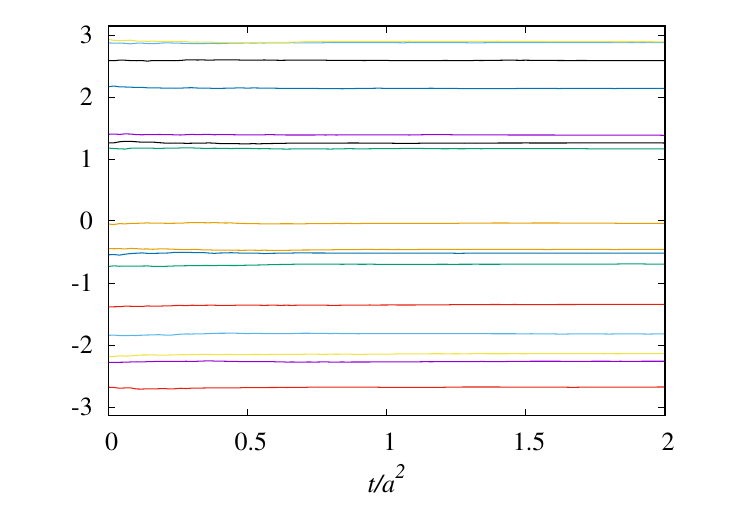}
\vspace{-11mm}
\end{center}
\caption{Phases of the monopole part of the averaged Polyakov loop on some configurations. The left and right panels are the results at $\beta=0.98$ and $1.10$, respectively.}
\label{fig:aplphasemo}
\end{figure}

\begin{figure}[tb]
\begin{center}
\vspace{0mm}
\includegraphics[width=8.1cm]{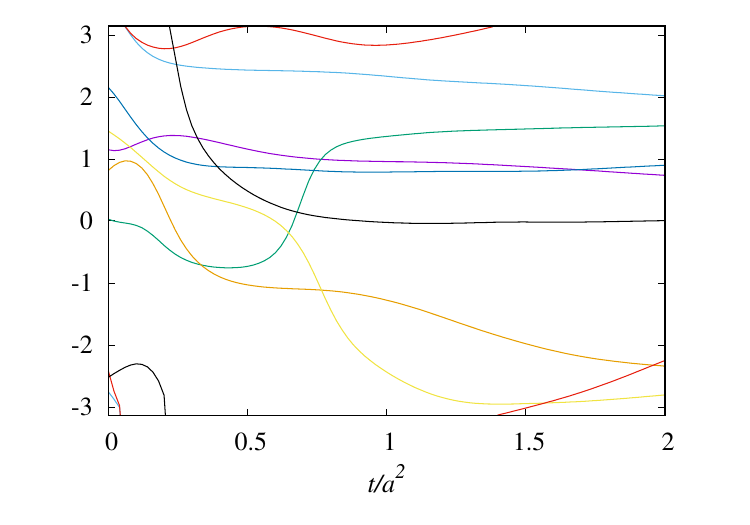}
\hspace{0mm}
\includegraphics[width=8.1cm]{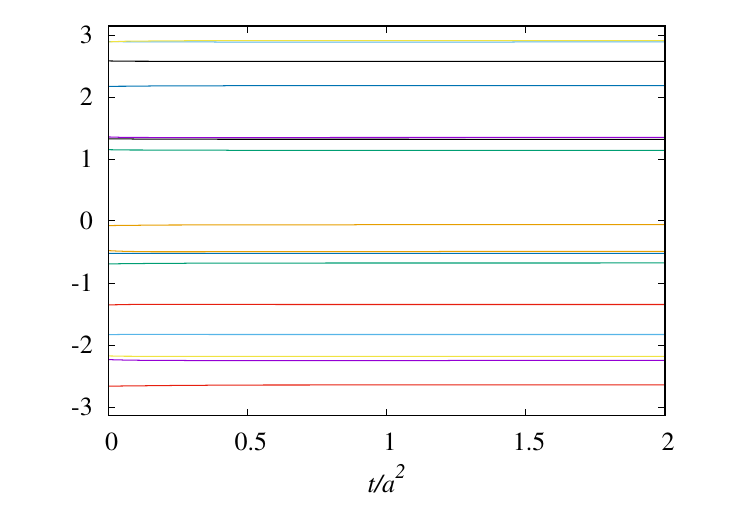}
\vspace{-11mm}
\end{center}
\caption{Phases of the averaged Polyakov loop on some configurations. The left and right panels are the results at $\beta=0.98$ and $1.10$, respectively.}
\label{fig:aplphase}
\end{figure}

Since the complex phase of the Polyakov loop from monopoles is given by the solid angle when looking at Dirac sheets surrounded by monopole currents, it is expected that the coarse graining of the gradient flow hardly changes the phase.
In this section, we study how the complex phases of the Polyakov loops change under the gradient flow.
Figure~\ref{fig:lplphasemo} shows the complex phase change of the local Polyakov loop from monopoles  $L_m^{\rm (loc)} (\vec{x})$ during the gradient flow. 
The left and right panels are the results at several points $\vec{x}$ on one configuration in the confinement phase $(\beta=0.98)$ and the deconfinement phase $(\beta=1.10)$, respectively. 
The lattice size is $32^3 \times 8$.
These are $\Omega (\vec{x})/2$ and the horizontal axis is the flow time $t/a^2$.
The monopole and the Dirac sheet are integer variables, the variation of $\Omega (\vec{x})$ under the gradient flow is discrete and hard to change.
Since the complex phase is uniformly distributed before the gradient flow in the confinement phase, the phase will be uniform after the gradient flow. Thus the gradient flow does not break the $U(1)$ symmetry.
In the deconfinement phase, the value of $\Omega (\vec{x})$ is almost the same at every point, which is given by the additional complex phase discussed so far, and the $\Omega (\vec{x})$ does not change by the gradient flow.
The histogram of half of the solid angle before the gradient flow is shown in Fig.~\ref{fig:sangpbc}, and the histogram after the gradient flow with $t/a^2=2.0$ is Fig.~\ref{fig:sangpbct2}.
The histogram does not change much by the gradient flow. 
The only change is the width of the peak in the deconfinement phase.

The complex phase of the local Polyakov loop at several points $\vec{x}$ on one configuration is plotted in Fig.~\ref{fig:lplphase} as a function of $t/a^2$.
Since the local Polyakov loop at $\vec{x}$ is given by $L_m^{\rm (loc)} (\vec{x}) \cdot L_f^{\rm (loc)} (\vec{x})$, the complex phase of the local Polyakov loop is the sum of the phases of the monopole part and the $F_{\mu \nu}$ part.
If the complex phase of the monopole part is uniformly distributed before the gradient flow in the confinement phase, the phase of the local Polyakov loop will be uniform after the gradient flow even when the phase from the $F_{\mu \nu}$ part is added.
The complex phase of the monopole part is almost the same at all $\vec{x}$ before the gradient flow in the deconfinement phase. Then, the phase of the local Polyakov loop becomes the phase of the monopole part $L_m^{\rm (loc)} (\vec{x})$ after the gradient flow because the $F_{\mu \nu}$ part $L_f^{\rm (loc)} (\vec{x})$ approaches one as the flow time increases.

Figure~\ref{fig:aplphasemo} is the result of the monopole part of the Polyakov loop $L_m$ averaged over the space. 
Each line is obtained on one configuration. 
As with the local Polyakov loop, the complex phase does not change much with the gradient flow.
The difference between the averaged Polyakov loop and the local Polyakov loop is that the additional complex phase in the deconfinement phase is different for each configuration.
The complex phase of $L_m$ is uniformly distributed and does not vary with the gradient flow.
The flow time dependence of the complex phase of the Polyakov loop on each configuration is shown in Fig.~\ref{fig:aplphase} for several configurations.
The difference from the monopole part is that the complex phase changes continuously, not discretely because the contribution from $F_{\mu \nu}$ is added.
Similar to the phase of the monopole part, the distribution of the phase of the Polyakov loop on each configuration remains uniform even after the gradient flow.
Thus, the $U(1)$ symmetry of the Polyakov loop is also preserved.

\section{Conclusions and outlook}
\label{conclusion}

We discussed the reason why the property of color confinement is not lost by the gradient flow even though the field strength is weakened.
In the gradient flow method, we solve a kind of diffusion equation and coarse grain the gauge field.
We expected that the confinement property is preserved because there is something in the background that is stable against coarse graining, such as topological quantities.
Performing Monte Carlo simulations, we investigated $U(1)$ lattice gauge theory, in which the cause of the confinement is believed to be condensation of magnetic monopoles.

We confirmed that the magnetic monopole does not disappear by the gradient flow in the confinement phase.
We usually use the gradient flow equation Eq.~(\ref{eq:coflow}), which considers the compactness of the gauge group.
However, we also investigated the case of the flow equation without considering compactness Eq.~(\ref{eq:noncoflow}).
We find that the compact flow equation keeps the confinement property, but the noncompact one breaks this property.
At the same time, the number of monopoles in the confinement phase does not decrease significantly for the compact flow equation, but decreases rapidly using the noncompact flow equation.

Wilson loops were calculated by decomposing them into contributions from field strength and monopoles.
The string tension is generated only from the monopole contribution, both before and after the flow.
We found that field strength does not contribute to the string tension, and that decreasing field strength due to the gradient flow does not affect the string tension.
The fact that the number of monopoles does not decrease in the gradient flow is strongly related to the fact that the string tension does not disappear.

The Polyakov loop can be also decomposed into contributions from field strength and monopoles, and we investigated how they change near the deconfinement phase transition.
Only the contribution from monopoles has the property of the order parameter of the center symmetry, which changes from zero to a finite value in the deconfinement phase transition; 
whereas, the contribution from the field strength does not change.

The monopole part of the Polyakov loop has the $U(1)$ center symmetry that is broken in the phase transition.
We discussed the relationship between monopoles and the center symmetry.
The unphysical complex phase by the center symmetry is created from Dirac strings which are also unphysical.
The gradient flow equation considering compactness does not break the center symmetry.
During the gradient flow, the center symmetry is maintained through the Dirac strings connected to the monopoles.

In $SU(3)$ gauge theory as well, it is important to perform a gradient flow that does not break the center symmetry in order not to lose the confinement property.
The gradient flow equation commonly used in lattice QCD takes into consideration the compactness of the gauge group and has the $Z_3$ center symmetry.
Since the center symmetry is important, it may be interesting to investigate the relationship with the $Z_3$ center vortices and so on \cite{Langfeld:1998cz,Greensite:2003bk}. 
If you want to discuss a magnetic monopole in QCD, you can think of a magnetic monopole of the $U(1)$ part extracted by the Abelian projection \cite{tHooft:1981bkw} as a straightforward extension.

\vspace{5mm}
\noindent\textbf{Acknowledgments}
%\subsection*{Acknowledgments} 

The authors thank the members of the WHOT-QCD Collaboration for useful discussions and comments.
This work was in part supported by JSPS KAKENHI Grant Numbers JP21K03550, JP20H01903, JP19H05146, and the HPCI System Research project (Project ID: hp220020, hp220024).

\appendix

\section{Volume dependence of the Creutz ratio}
\label{sec:volume}

\begin{figure}[tb]
\begin{center}
\vspace{0mm}
\includegraphics[width=8.1cm]{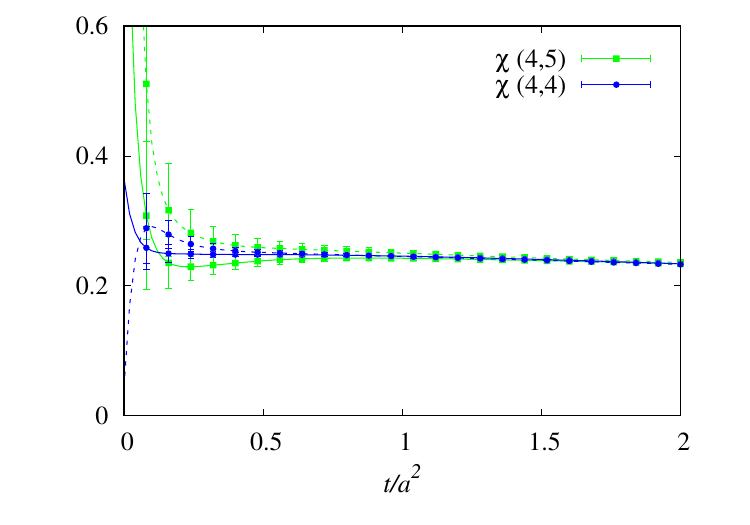}
\hspace{0mm}
\includegraphics[width=8.1cm]{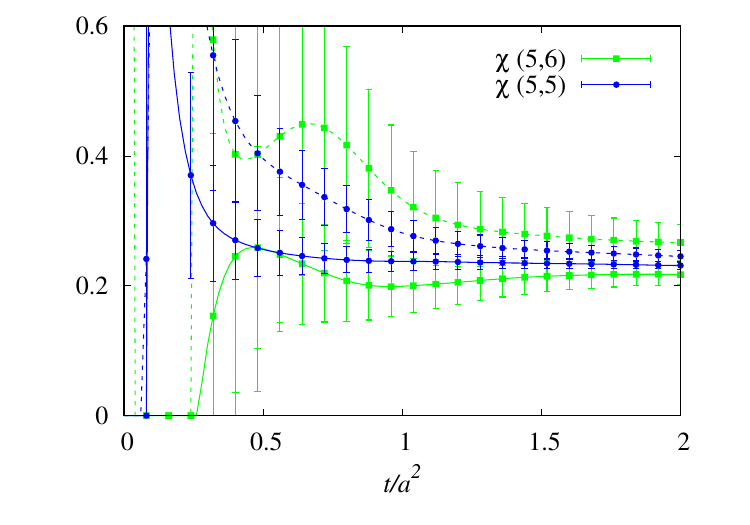}
\vspace{-11mm}
\end{center}
\caption{Creutz ratios $\chi(i,j)$ as functions of the flow time measured at $\beta=0.99$ on $20^4$ (solid lines) and $16^4$ (dashed lines) lattices. 
}
\label{fig:chiapbc}
\end{figure}

To investigate the finite volume effect in the Creutz ratio, we perform simulations at $\beta=0.99$ on lattices with $N_s^3 \times N_t = 16^4$ and $20^4$.
The number of independent configurations is 10000.
The results of the Creutz ratios $\chi (i,j)$ are plotted in Fig.~\ref{fig:chiapbc}.
The solid lines and dashed lines are the result on $20^4$ and $16^4$ lattices, respectively.
The left panel is $\chi (4,4)$ (blue) and $\chi (4,5)$ (green) and the right panel is $\chi (5,5)$ (blue) and $\chi (5,6)$ (green).
The results of $\chi (4,4)$ and $\chi (4,5)$ on the $16^4$ and $20^4$ lattices are consistent within statistical error.
However, volume dependence is seen in $\chi (5,5)$ and $\chi (5,6)$ on $16^4$ lattice and $\chi (5,6)$ on $20^4$ lattice.
These are different from $\chi (4,4)$ and $\chi (4,5)$. 
This suggests that a finite volume effect is visible in Creutz ratios containing Wilson loops with a side length greater than $N_s/3$ for a $N_s^4$ lattice.
Since $\chi (i,j)$ is given by $W(i,j)$, $W(i+1,j)$, and $W(i+1, j+1)$, for the $20^4$ lattice,
$\chi (i,j)$ is affected by the finite volume effect for $i, j \ge 6$.
Therefore, the results of the Creutz ratios with the side lengths less than 6 are shown in Figs.~\ref{fig:creutz-t}, \ref{fig:cremonflow}, \ref{fig:crephflow}, and 
Table~\ref{tab:chi}.

\section{Monte Carlo Simulation with antiperiodic boundary conditions}
\label{sec:antiprio}

The Coulomb propagator cannot be strictly defined under periodic boundary conditions.
This issue confuses us in discussing the relationship between Polyakov loop and the monopole condensation in Sec.~\ref{sec:transition}.
Therefore, we change the boundary conditions of the Coulomb propagator so that the definition,
$\partial'_{\mu} \partial_{\mu} D(x) = - \delta_{x,0}$, 
can be strictly satisfied.
We impose antiperiodic boundary conditions on $D(x)$ in the spatial directions.
The Coulomb propagator can be defined exactly if at least one direction is an antiperiodic boundary condition, since there is no constant mode in $D(x)$.

In determining the boundary conditions, it is important to cancel the surface term at the boundary coming from the total derivative term in integration by parts.
Integration by parts has been done several times to derive the monopole part of the Polyakov loop Eq.~(\ref{eq:p3mon}).
Considering that the surface term is always in the combination of $D(x-x') \, \theta_{\mu}$ or $D(x-x') \, k_{\mu}$, all of $D(x)$, $\theta_{\mu} (x)$ and $k_{\mu}(x)$ must be antiperiodic boundary conditions:
\begin{eqnarray}
D(x) = -D(x+ N_{s} \hat{i}), \hspace{5mm}
\theta_{\mu}(x) = -\theta_{\mu}(x+ N_{s} \hat{i}), \hspace{5mm}
k_{\mu}(x) = - k_{\mu}(x+ N_{s} \hat{i}) , 
\end{eqnarray}
where $\hat{i}$ means the next site in the $i$ direction.
Then, the surface terms are canceled.
However, due to the finite temperature system, the gauge field in the time direction must impose periodic boundaries.
Therefore, only the boundary conditions in the spatial directions are set to be antiperiodic.
The link fields, $U_{\mu}(x)=e^{i \theta_{\mu}(x)}$, beyond the boundary are the complex conjugate,
\begin{eqnarray}
U_{\mu} (x) = U^{\ast}_{\mu} (x + N_{s} \hat{i}). 
\end{eqnarray}

For the case of antiperiodic boundary condition, the action of the gauge field is not symmetric under the center transformation, $U_4(x) \to e^{i \phi} U_4(x)$, at the boundaries.
The cross-boundary plaquette changes as follows: 
\begin{eqnarray}
&& \hspace{-10mm}
U_{\mu}(x) U_{4}^{*} (x+(1-N_s) \hat{\mu}) U^{*}_{\mu}(x+\hat{4}) U^{*}_{4}(x) \nonumber \\
&\to& U_{\mu}(x) \{ e^{i \phi} U_{4}(x+(1-N_s) \hat{\mu}) \}^{*}
U^{*}_{\mu}(x+\hat{4}) \{ e^{i \phi} U_4(x) \}^{*} \nonumber \\
&=&  e^{-2i \phi} U_{\mu}(x) U_{4}^{*}(x+(1-N_s) \hat{\mu})
U^{*}_{\mu}(x+\hat{4}) U^{*}_{4}(x) .
\end{eqnarray}
Thus, except in the case of $\phi = \pm \pi$, the symmetry is broken.
Therefore, the center symmetry is reduced from $U(1)$ symmetry to $Z_2$ symmetry.

\begin{figure}[tb]
\begin{center}
\vspace{0mm}
\includegraphics[width=8.1cm]{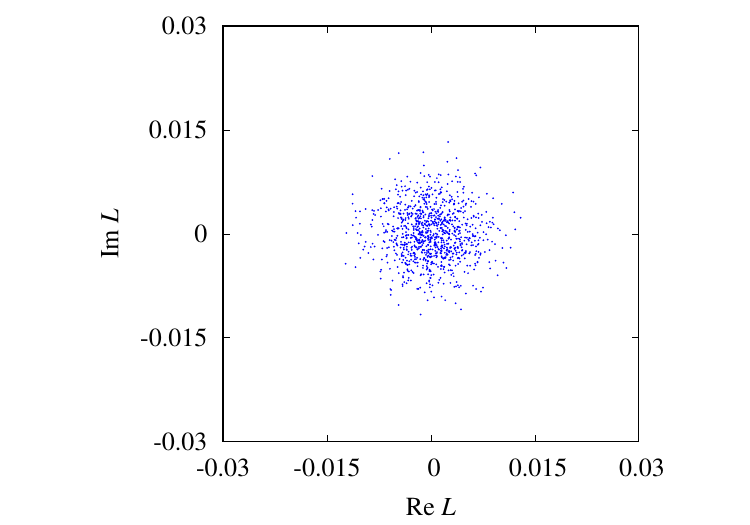}
\hspace{0mm}
\includegraphics[width=8.1cm]{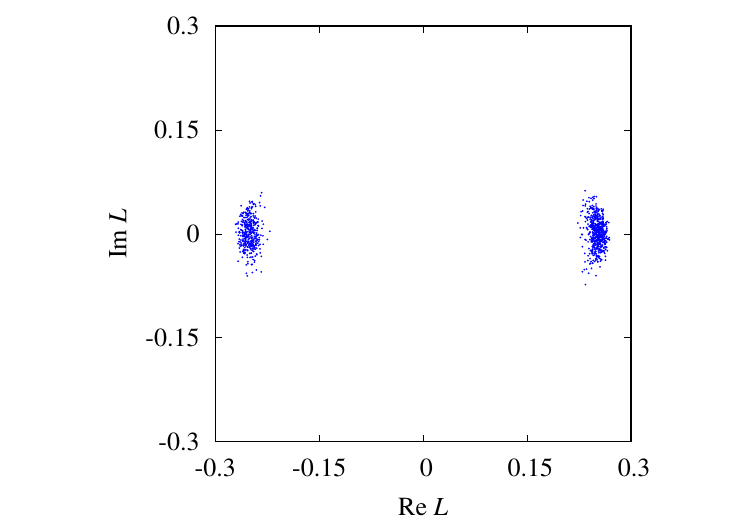}
\vspace{-11mm}
\end{center}
\caption{Distribution of the Polyakov loop in the complex plane with antiperiodic boundary condition.  
The left and right panels are the results for the confinement phase $(\beta=0.90)$ and the deconfinement phase $(\beta=1.10)$ on a $32^3 \times 8$ lattice, respectively.
}
\label{fig:plapbc}
\end{figure}

\begin{figure}[tb]
\begin{center}
\vspace{0mm}
\includegraphics[width=8.1cm]{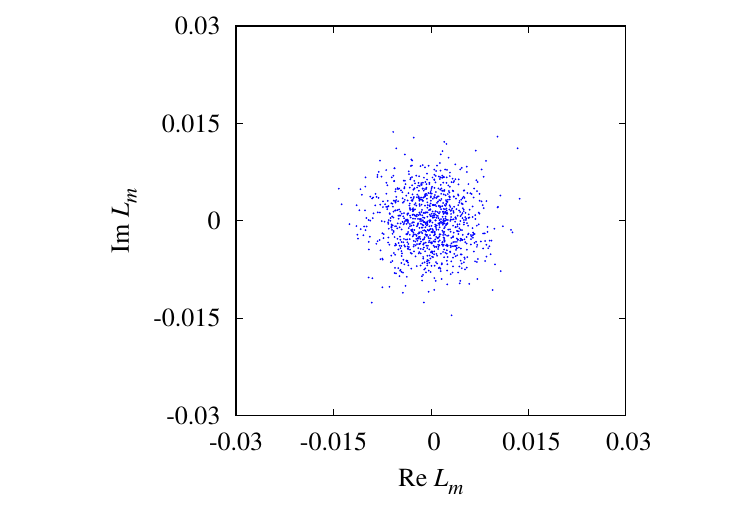}
\hspace{0mm}
\includegraphics[width=8.1cm]{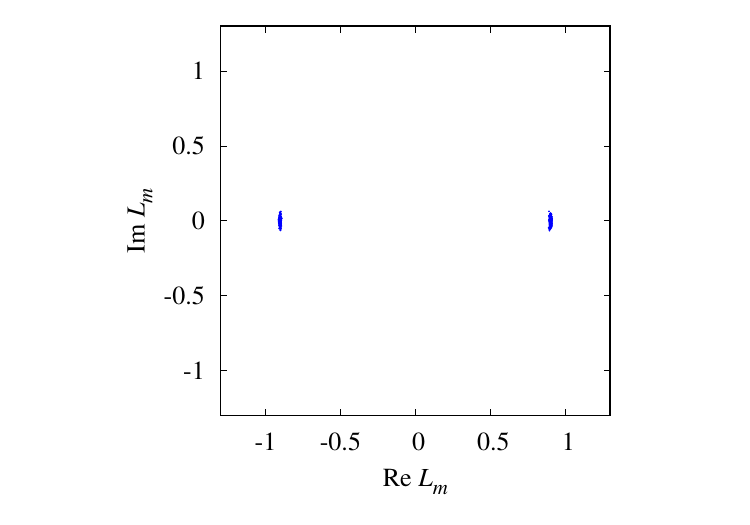}
\vspace{-11mm}
\end{center}
\caption{Distribution of the monopole contribution to the Polyakov loop in the complex plane  with antiperiodic boundary condition.  
The left and right panels are the results at $\beta=0.90$ and $1.10$ on a $32^3 \times 8$ lattice, respectively.
}
\label{fig:plmonapbc}
\end{figure}

Monte Carlo simulations with the antiperiodic boundary condition are performed.
In Fig.~\ref{fig:plapbc}, we plot the distribution of the Polyakov loop on each configuration.
The lattice size is $32^3 \times 8$.
The number of configurations is 1000 for each $\beta$.
The left panel is $\beta=0.90$ in the confinement phase, and the right panel is $\beta=1.10$ in the deconfinement phase.
The distribution of the Polyakov loop is $Z_2$ symmetric from $L$ to $-L$, corresponding to the symmetry change from $U(1)$ to $Z_2$.
The contribution of the Polyakov loop from the monopole is also calculated and shown in Fig.~\ref{fig:plmonapbc}.
The distribution of the monopole contribution is also $Z_2$ symmetric from $L_m$ to $-L_m$.

We have discussed in Sec.~\ref{sec:center} that the complex phase of the Polyakov loop from monopoles is created from the Dirac sheet bubbles for periodic boundary conditions.
However, if the Coulomb propagator can be defined strictly, the behavior of the Polyakov loop can be understood from the solid angle of the monopole in Eq.~(\ref{eq:psang}).
Then, the Dirac sheet without monopoles does not affect the Polyakov loop, and there is no reason to change the complex phase by the Dirac sheet.
Only the sign of the Polyakov loop changes with the infinitely wide Dirac sheet.
The sign changing without monopoles corresponds to the center transformation reduced from $U(1)$ to $Z_2$.

As shown in Fig.~\ref{fig:plmonapbc} for antiperiodic boundary conditions, the Polyakov loop values are distributed near the origin in the confinement phase and around two points on the real axis in the deconfinement phase.
This result can be understood from the solid angle of the monopoles.
The probability distribution of the solid angle is flat in the confinement phase because the monopoles are distributed throughout the space.
Then, the Polyakov loop on each configuration becomes zero when the spatial average is taken.
On the other hand, in the deconfinement phase, the solid angle is distributed near zero because there are very few monopoles.
Then, the spatial average of the Polyakov loop is a nonzero value near the real axis, and the distribution is $Z_2$ symmetric due to the infinitely wide Dirac sheet.


\begin{thebibliography}{99}

%\cite{Narayanan:2006rf}
\bibitem{Narayanan:2006rf}
R.~Narayanan and H.~Neuberger,
Infinite N phase transitions in continuum Wilson loop operators,
J. High Energy Phys. \textbf{03}, 064 (2006).
%doi:10.1088/1126-6708/2006/03/064
%[arXiv:hep-th/0601210 [hep-th]].

%\cite{Luscher:2009eq}
\bibitem{Luscher:2009eq}
M.~Luscher,
Trivializing maps, the Wilson flow and the HMC algorithm,
Commun. Math. Phys. \textbf{293}, 899 (2010).
%doi:10.1007/s00220-009-0953-7
%[arXiv:0907.5491 [hep-lat]].

%\cite{Luscher:2010iy}
\bibitem{Luscher:2010iy}
M.~L\"uscher,
Properties and uses of the Wilson flow in lattice QCD,
J. High Energy Phys. \textbf{08}, 071 (2010).
%[erratum: JHEP \textbf{03}, 092 (2014)]
%doi:10.1007/JHEP08(2010)071
%[arXiv:1006.4518 [hep-lat]].

%\cite{Suzuki:2013gza}
\bibitem{Suzuki:2013gza}
H.~Suzuki,
Energy\textendash{}momentum tensor from the Yang\textendash{}Mills gradient flow,
Prog. Theor. Exp. Phys. \textbf{2013}, 083B03 (2013).
%[erratum: PTEP \textbf{2015}, 079201 (2015)]
%doi:10.1093/ptep/ptt059
%[arXiv:1304.0533 [hep-lat]].

%\cite{Makino:2014taa}
\bibitem{Makino:2014taa}
H.~Makino and H.~Suzuki,
Lattice energy\textendash{}momentum tensor from the Yang\textendash{}Mills gradient flow\textemdash{}inclusion of fermion fields,
Prog. Theor. Exp. Phys. \textbf{2014}, 063B02 (2014).
%[erratum: PTEP \textbf{2015}, 079202 (2015)]
%doi:10.1093/ptep/ptu070
%[arXiv:1403.4772 [hep-lat]].

%\cite{Asakawa:2013laa}
\bibitem{Asakawa:2013laa}
M.~Asakawa, T.~Hatsuda, E.~Itou, M.~Kitazawa, H.~Suzuki [FlowQCD],
Thermodynamics of SU(3) gauge theory from gradient flow on the lattice,
Phys. Rev. D \textbf{90}, 011501(R) (2014).
%[erratum: Phys. Rev. D \textbf{92}, no.5, 059902 (2015)]
%doi:10.1103/PhysRevD.90.011501
%[arXiv:1312.7492 [hep-lat]].

%\cite{Kitazawa:2016dsl}
\bibitem{Kitazawa:2016dsl}
M.~Kitazawa, T.~Iritani, M.~Asakawa, T.~Hatsuda and H.~Suzuki,
Equation of State for SU(3) Gauge Theory via the Energy-Momentum Tensor under Gradient Flow,
Phys. Rev. D \textbf{94}, 114512 (2016).
%doi:10.1103/PhysRevD.94.114512
%[arXiv:1610.07810 [hep-lat]].

%\cite{Iritani:2018idk}
\bibitem{Iritani:2018idk}
T.~Iritani, M.~Kitazawa, H.~Suzuki and H.~Takaura,
Thermodynamics in quenched QCD: energy\textendash{}momentum tensor with two-loop order coefficients in the gradient-flow formalism, 
Prog. Theor. Exp. Phys. \textbf{2019}, 023B02 (2019).
%doi:10.1093/ptep/ptz001
%[arXiv:1812.06444 [hep-lat]].

%\cite{Taniguchi:2016ofw}
\bibitem{Taniguchi:2016ofw}
Y.~Taniguchi, S.~Ejiri, R.~Iwami, K.~Kanaya, M.~Kitazawa, H.~Suzuki, T.~Umeda and N.~Wakabayashi,
Exploring $N_{f}$ = 2+1 QCD thermodynamics from the gradient flow,
Phys. Rev. D \textbf{96}, 014509 (2017).
%[erratum: Phys. Rev. D \textbf{99}, no.5, 059904 (2019)]
%doi:10.1103/PhysRevD.96.014509
%[arXiv:1609.01417 [hep-lat]].

%\cite{Taniguchi:2016tjc}
\bibitem{Taniguchi:2016tjc}
Y.~Taniguchi, K.~Kanaya, H.~Suzuki and T.~Umeda,
Topological susceptibility in finite temperature ( 2+1 )-flavor QCD using gradient flow,
Phys. Rev. D \textbf{95}, 054502 (2017).
%doi:10.1103/PhysRevD.95.054502
%[arXiv:1611.02411 [hep-lat]].

%\cite{Taniguchi:2020mgg}
\bibitem{Taniguchi:2020mgg}
Y.~Taniguchi, S.~Ejiri, K.~Kanaya, M.~Kitazawa, H.~Suzuki, T.~Umeda [WHOT-QCD],
$N_f$ = 2+1 QCD thermodynamics with gradient flow using two-loop matching coefficients,
Phys. Rev. D \textbf{102}, 014510 (2020).
%[erratum: Phys. Rev. D \textbf{102}, no.5, 059903 (2020)]
%doi:10.1103/PhysRevD.102.014510
%[arXiv:2005.00251 [hep-lat]].

%\cite{Shirogane:2020muc}
\bibitem{Shirogane:2020muc}
M.~Shirogane, S.~Ejiri, R.~Iwami, K.~Kanaya, M.~Kitazawa, H.~Suzuki, Y.~Taniguchi, T.~Umeda [WHOT-QCD],
Latent heat and pressure gap at the first-order deconfining phase transition of SU(3) Yang-Mills theory using the small flow-time expansion method,
Prog. Theor. Exp. Phys. \textbf{2021}, 013B08 (2021).
%doi:10.1093/ptep/ptaa184
%[arXiv:2011.10292 [hep-lat]].

%\cite{Banks:1977cc}
\bibitem{Banks:1977cc}
T.~Banks, R.~Myerson and J.~B.~Kogut,
Phase Transitions in Abelian Lattice Gauge Theories,
Nucl. Phys. B \textbf{129}, 493 (1977).
%doi:10.1016/0550-3213(77)90129-8

\bibitem{Shuryak2021}
E.~Shuryak,
Nonperturbative Topological Phenomena in QCD and Related Theories,
(Springer, 2021).

%\cite{tHooft:1981bkw}
\bibitem{tHooft:1981bkw}
G.~'t Hooft,
Topology of the Gauge Condition and New Confinement Phases in Nonabelian Gauge Theories,
Nucl. Phys. B \textbf{190}, 455 (1981).
%doi:10.1016/0550-3213(81)90442-9

%\cite{Kronfeld:1987ri}
\bibitem{Kronfeld:1987ri}
A.~S.~Kronfeld, M.~L.~Laursen, G.~Schierholz and U.~J.~Wiese,
Monopole Condensation and Color Confinement,
Phys. Lett. B \textbf{198}, 516 (1987).
%doi:10.1016/0370-2693(87)90910-5

%\cite{Kronfeld:1987vd}
\bibitem{Kronfeld:1987vd}
A.~S.~Kronfeld, G.~Schierholz and U.~J.~Wiese,
Topology and Dynamics of the Confinement Mechanism,
Nucl. Phys. B \textbf{293}, 461 (1987).
%doi:10.1016/0550-3213(87)90080-0

%\cite{Shiba:1994ab}
\bibitem{Shiba:1994ab}
H.~Shiba and T.~Suzuki,
Monopoles and string tension in SU(2) QCD,
Phys. Lett. B \textbf{333}, 461 (1994).
%doi:10.1016/0370-2693(94)90168-6
%[arXiv:hep-lat/9404015 [hep-lat]].

%\cite{Ejiri:1994uw}
\bibitem{Ejiri:1994uw}
S.~Ejiri, S.~i.~Kitahara, Y.~Matsubara and T.~Suzuki,
String tension and monopoles in $T \neq 0$ SU(2) QCD,
Phys. Lett. B \textbf{343}, 304 (1995).
%doi:10.1016/0370-2693(94)01457-N
%[arXiv:hep-lat/9407022 [hep-lat]].

%\cite{Suzuki:1994ay}
\bibitem{Suzuki:1994ay}
T.~Suzuki, S.~Ilyar, Y.~Matsubara, T.~Okude and K.~Yotsuji,
Polyakov loops and monopoles in QCD,
Phys. Lett. B \textbf{347}, 375 (1995).
%[erratum: Phys. Lett. B \textbf{351}, 603 (1995)]
%doi:10.1016/0370-2693(95)00068-V
%[arXiv:hep-lat/9408003 [hep-lat]].
%\cite{Ejiri:1995gd}

\bibitem{Ejiri:1995gd}
S.~Ejiri,
Monopoles and spatial string tension in the high temperature phase of SU(2) QCD,
Phys. Lett. B \textbf{376}, 163 (1996).
%doi:10.1016/0370-2693(96)00245-6
%[arXiv:hep-lat/9510027 [hep-lat]].

%\cite{DAlessandro:2007lae}
\bibitem{DAlessandro:2007lae}
A.~D'Alessandro and M.~D'Elia,
Magnetic monopoles in the high temperature phase of Yang-Mills theories,
Nucl. Phys. B \textbf{799}, 241 (2008).
%doi:10.1016/j.nuclphysb.2008.03.002
%[arXiv:0711.1266 [hep-lat]].

%\cite{Suzuki:2009xy}
\bibitem{Suzuki:2009xy}
T.~Suzuki, M.~Hasegawa, K.~Ishiguro, Y.~Koma and T.~Sekido,
Gauge invariance of color confinement due to the dual Meissner effect caused by Abelian monopoles,
Phys. Rev. D \textbf{80}, 054504 (2009).
%doi:10.1103/PhysRevD.80.054504
%[arXiv:0907.0583 [hep-lat]].

%\cite{Creutz:1983ev}
\bibitem{Creutz:1983ev}
M.~Creutz, L.~Jacobs and C.~Rebbi,
Monte Carlo Computations in Lattice Gauge Theories,
Phys. Rept. \textbf{95}, 201 (1983).
%doi:10.1016/0370-1573(83)90016-9

%\cite{DeGrand:1980eq}
\bibitem{DeGrand:1980eq}
T.~A.~DeGrand and D.~Toussaint,
Topological Excitations and Monte Carlo Simulation of Abelian Gauge Theory,
Phys. Rev. D \textbf{22}, 2478 (1980).
%doi:10.1103/PhysRevD.22.2478

%\cite{Stack:1991zp}
\bibitem{Stack:1991zp}
J.~D.~Stack and R.~J.~Wensley,
Monopoles, quark confinement and screening in four-dimensional U(1) lattice gauge theory,
Nucl. Phys. B \textbf{371}, 597 (1992).
%doi:10.1016/0550-3213(92)90688-8

%\cite{Ejiri:1996sz}
\bibitem{Ejiri:1996sz}
S.~Ejiri,
Monopole condensation and Polyakov loop in finite temperature pure QCD,
Nucl. Phys. B Proc. Suppl. \textbf{53}, 491 (1997).
%doi:10.1016/S0920-5632(96)00696-2
%[arXiv:hep-lat/9608001 [hep-lat]].

%\cite{Ejiri:1998xf}
\bibitem{Ejiri:1998xf}
S.~Ejiri,
Monopole condensation and quark confinement at finite temperature QCD,
Nucl. Phys. A \textbf{629}, 89C (1998).
%doi:10.1016/S0375-9474(97)00670-2

%\cite{Langfeld:1998cz}
\bibitem{Langfeld:1998cz}
K.~Langfeld, O.~Tennert, M.~Engelhardt and H.~Reinhardt,
Center vortices of Yang-Mills theory at finite temperatures,
Phys. Lett. B \textbf{452}, 301 (1999).
%doi:10.1016/S0370-2693(99)00252-X
%[arXiv:hep-lat/9805002 [hep-lat]].

%\cite{Greensite:2003bk}
\bibitem{Greensite:2003bk}
J.~Greensite,
The Confinement problem in lattice gauge theory,
Prog. Part. Nucl. Phys. \textbf{51}, 1 (2003).
%doi:10.1016/S0146-6410(03)90012-3
%[arXiv:hep-lat/0301023 [hep-lat]].


\end{thebibliography}
\end{document}